\documentclass{emulateapj}

\usepackage{natbib} 
\usepackage{epsf}
\usepackage{psfig}
\usepackage{lscape}
\usepackage{graphicx}

\begin{document}

\def\mpch {$h^{-1}$ Mpc} 
\def\kpch {$h^{-1}$ kpc} 
\def\kms {km s$^{-1}$} 
\def\lcdm {$\Lambda$CDM } 
\def\etal {et al.}
\def\dzz {\sigma_z/(1+z)}
\def\ddzz {\Delta_z/(1+z)}
\def\hmpcC {h^{-3}\rm\,Mpc^3}
\def\ihmpcC {h^{3}\rm\,Mpc^{-3}}
\def\micron {\mbox{$\mu$m}}

\title{The PRIsm MUlti-Object Survey (PRIMUS) I: Survey Overview and Characteristics}
\author{
Alison L. Coil\altaffilmark{1,2}, 
Michael R. Blanton\altaffilmark{3}, 
Scott M. Burles\altaffilmark{4}, 
Richard J. Cool\altaffilmark{5,6}, 
Daniel J. Eisenstein\altaffilmark{7}, 
John Moustakas\altaffilmark{1}, 
Kenneth C. Wong\altaffilmark{7}, 
Guangtun Zhu\altaffilmark{3},
James Aird\altaffilmark{1},
Rebecca A. Bernstein\altaffilmark{8},
Adam S. Bolton\altaffilmark{9},
David W. Hogg\altaffilmark{3}}

\altaffiltext{1}{Department of Physics, Center for Astrophysics and Space Sciences, University of California, 9500 Gilman Dr., La Jolla, San Diego, CA 92093}
\altaffiltext{2}{Alfred P. Sloan Foundation Fellow}
\altaffiltext{3}{Center for Cosmology and Particle Physics, Department of Physics, New York University, 4 Washington Place, New York, NY 10003}
\altaffiltext{4}{D.E. Shaw \& Co., L.P., 20400 Stevens Creek Blvd., Suite 850, Cupertino, CA 95014}
\altaffiltext{5}{Department of Astrophysical Sciences, Princeton University, Peyton Hall, Princeton, NJ 08544}
\altaffiltext{6}{Hubble Fellow, Princeton-Carnegie Fellow}
\altaffiltext{7}{Steward Observatory, The University of Arizona, 933 N. Cherry Ave., Tucson, AZ 85721}
\altaffiltext{8}{Department of Astronomy and Astrophysics, UCO/Lick Observatory, University of California, 1156 High Street, Santa Cruz, CA 95064}
\altaffiltext{9}{Department of Physics and Astronomy, University of Utah, Salt Lake City, UT 84112}

\begin{abstract}
We present the PRIsm MUlti-object Survey (PRIMUS), a spectroscopic 
faint galaxy redshift survey to $z\sim1$. 
PRIMUS uses a low-dispersion prism and slitmasks to observe
$\sim$2,500 objects at once in a 0.18 deg$^2$ field of view, using the 
IMACS camera on the Magellan I Baade 6.5m telescope at Las Campanas 
Observatory. PRIMUS covers a total of 9.1 deg$^2$ of sky to a depth of
$i_{\rm AB}\sim23.5$ in seven different deep, multi-wavelength fields that have 
coverage from {\it GALEX, Spitzer} and either {\it XMM} or {\it Chandra}, 
as well as multiple-band optical and near-IR coverage.  PRIMUS includes 
$\sim$130,000 robust redshifts of unique objects with a redshift precision of
$\dzz\sim0.005$.  The redshift distribution peaks at $z\sim0.6$ 
and extends to $z=1.2$ for galaxies and $z=5$ for broad-line AGN.
The motivation, observational 
techniques, fields, target selection, slitmask design, and observations 
are presented here, with a brief summary of the redshift precision; a 
companion paper presents the data reduction, redshift 
fitting, redshift confidence, and survey completeness.  PRIMUS is the 
largest faint galaxy survey undertaken to date.  The high targeting 
fraction ($\sim80$\%) and large survey size will allow for precise 
measures of galaxy properties and large-scale structure to $z\sim1$.
\end{abstract}


\section{Introduction} \label{sec:introduction}

Redshift surveys are a crucial tool for understanding galaxy
evolution.  As pioneering large redshift surveys of the local Universe 
dramatically demonstrated \citep[e.g., the CfA survey,][]{Davis82},
redshifts remove the effects of projection, reveal the cosmic web, and
allow conversion of observed fluxes and sizes to physical restframe
quantities. 
Armed with redshifts, one can determine key statistical
properties of galaxy populations such as luminosity functions, 
merger rates, and correlation functions, across cosmic time.  Combining
redshifts with multi-wavelength imaging from the ultraviolet (UV) to the 
infrared (IR) allows one to further measure star formation rates (SFRs)
 and stellar masses, while X-ray data allow studies 
of AGN properties.

At low redshift, the largest galaxy redshift surveys are the Sloan
Digital Sky Survey \citep[SDSS,][]{York00}, the Two Degree Field
Galaxy Redshift Survey \citep[2dFGRS,][]{Colless01}, and the Six Degree Field
Galaxy Redshift Survey \citep[6dFGS,][]{Jones09}.  These surveys
extend to $z\sim0.2$ and contain hundreds of thousands of galaxy
redshifts, mapping out volumes of $\sim10^{7.6} \ h^3$ Mpc$^{-3}$ and
$\sim10^{8.2} \ h^3$ Mpc$^{-3}$ for 2dFGRS and SDSS, respectively.
Enormous datasets of this size are required to both overcome the
cosmic variance of large-scale clustering and to measure joint
distributions of properties such as color and luminosity across a
range of environments.  It is only through studying joint properties
of galaxies as well as their full distribution functions, spanning a
wide range of redshifts, that the evolution of galaxies will be understood.

Several flux-limited redshift surveys have begun to probe initial
volumes of the distant universe. At intermediate redshift, the
AGN and Galaxy Evolution Survey (AGES, Kochanek et al. in preparation)
is a wide-field survey that has
obtained redshifts for $\sim$13,000 galaxies over $\sim$8 deg$^2$
to $I_{AB}=20.4$.  The median redshift of the AGES
galaxy sample is $z\sim0.3$.  Deeper surveys have targeted the more
distant universe; the largest of these include COMBO-17 \citep{Wolf03}, 
DEEP2 \citep[][Faber in preparation]{Davis03}, 
the VIMOS VLT Deep Survey \citep[VVDS,][]{Lefevre05, Garilli08}, 
and zCOSMOS \citep{Lilly07}.  Of
these, the largest area surveyed is the VVDS-Wide survey, which has
covered $\sim$6 deg$^2$ to a depth of $I_{AB}=22.5$; the other
surveys cover half this area, at most.  The deepest of these 
surveys are DEEP2,
which covered $\sim$3 deg$^2$ to a depth of $R_{AB}=24.1$, and
VVDS-Deep, which covered $\sim$0.6 deg$^2$ to a depth of
$I_{AB}=24$.  The largest samples in terms of number of redshifts 
are the DEEP2 and VVDS-Wide surveys,
with $\sim$30,000 and $\sim$20,000 current secure redshifts each.  The median
redshift of these samples is $\sim z=0.5-1$.  Surveys at higher
redshift ($z>1.5$) have not been flux-limited in a single photometric band, 
but rather use various
color-selection techniques to identify galaxy candidates at specific
redshifts \citep[e.g.,][]{Steidel04,Quadri07, vanDokkum09}, 
resulting in samples of typically a few thousand objects.

What is currently known about galaxy evolution results from 
surveys that extend to high redshift; however, the
sample sizes and volumes mapped by these surveys are still relatively
small compared to low redshift samples.  The samples are not large
enough to measure, for example, the luminosity function of blue and
red galaxies in multiple bins in redshift and environment.  The
dominant error in most statistical studies at $z>0.2$ is cosmic
variance\footnote[1]{E.g., in the COSMOS field, 1.4 deg. on a side, the 
fractional error in a count of an unbiased tracer of mass is 9\% at $z=1$ 
in a $\delta z=0.2$ bin.}; 
the volumes that have been surveyed are not yet large enough
to fully sample the overdensity distribution at a given redshift.  An
additional complication in trying to measure the evolution of a given
galaxy statistic is that there is often a gap in redshift between low
redshift samples at $z<0.2$ and higher redshift samples at $z\sim1$.
The deeper surveys do not have sufficiently large numbers of sources
or survey sufficient volumes at intermediate redshifts of $0.2<z<0.5$ to
continuously track a statistic over cosmic time with small error bars.  
While the AGES survey addresses this gap to some degree, it is not 
sufficiently deep to overlap the higher redshift surveys substantially 
in redshift.  
What is needed are significantly larger
surveys at intermediate redshift that are 1) large enough to have comparable
errors to local surveys such as 2dFGRS and SDSS, 2) deep enough to sample
typical $L_\ast$ galaxies at intermediate redshift, and 3) 
have the requisite multi-wavelength imaging needed to
 measure SFRs and stellar masses.

The deep, multi-wavelength imaging extending beyond the optical
regime that is required to carry out such a 
survey at intermediate redshifts is now available.
Simultaneous with the spectroscopic advances that have allowed the 
first redshift surveys to $z\sim1$ to be carried out, astronomy
has entered a golden age in wide-field imaging, both 
to survey local galaxies and those at $z>1$.
Modern imaging surveys have mapped a large volume ($\sim10^8 \ \hmpcC$) at
redshift $z\sim0$ in the ultraviolet with the Galaxy Evolution
Explorer \citep[{\it GALEX},][]{Martin05}, in the optical with SDSS, and in
the near-IR with the 2-Micron All Sky Survey
\citep[2MASS,][]{skrutskie97}.  The similarly ambitious United Kingdom
Infrared Deep Sky Survey \citep[UKIDDS,][]{Lawrence07} near-IR survey
is conducting a survey of 7500 deg$^2$ to $K=18.3$, 
and the Wide-field Infrared Survey Explorer
\citep[WISE,][]{Liu08}, which will survey the entire sky in the 
mid-IR ($3-25\mu$m), launched in late 2009. Even
more remarkable is that a similar volume has recently become available
at $z\sim1$.  
The UKIDDS deep survey to $K=21$, 
the {\it Spitzer} Wide-area Infrared Extragalactic Survey
\citep[SWIRE,][]{Lonsdale03}, the {\it Spitzer} Deep, Wide-Field Survey
\citep[SDWFS][]{Ashby09} and {\it Spitzer} GTO shallow program
\citep{Eisenhardt04}, 
the {\it GALEX} Deep Imaging Survey (DIS), and
companion deep optical imaging are covering about 70 deg$^2$ 
to depths that reach below $L_\ast$ at $z=1$ from the UV to the
far-IR.  

These deep imaging surveys can be harnessed
to study galaxy evolution using volumes sufficient to overcome
cosmic variance, but only by coupling them to large redshift
surveys.  Only with redshifts can restframe properties such as
luminosities, colors, SFRs, and stellar masses be robustly measured. 
The goal of the PRIsm MUlti-object Survey (PRIMUS), which we present
here, is to provide the spectroscopy necessary to make this science
possible. PRIMUS is the largest faint galaxy redshift survey performed
to date. PRIMUS has targeted multiple fields with existing deep
optical imaging, UV {\it GALEX} imaging, IR {\it Spitzer}/IRAC and MIPS imaging,
and X-ray {\it Chandra} and/or {\it XMM} coverage in order to provide the data 
to address these
issues.  By obtaining low-resolution spectra with a slitmask and
prism, we have assembled a unique and unprecedentedly large data set
containing $\sim$120,000 redshifts to $z<1.2$ (and $\sim$9,000 stars), covering 
over 9.1 deg$^2$ of sky with complementary multi-wavelength
imaging.  Using the IMACS instrument on the Magellan telescope,
which has a wide 0.2 deg$^2$ field, with a prism we observed
$\sim$2,500 targets simultaneously to depths of $i\sim23$ in one hour. We
observed $\sim$10,000 target galaxies per night, substantially 
larger than traditional spectroscopic surveys undertaken at Keck and
VLT.  Our survey has a redshift precision of $\dzz=0.5$\%, which is the
precision required to measure galaxy environments and clustering
statistics. This survey is approximately four times larger than DEEP2 and covers
one half the volume of 2dFGRS, but at a mean redshift of $z=0.56$ (for
the joint sample of galaxies and broad-line AGN).

This paper describes the design, survey characteristics, and data taken
for 
PRIMUS, which is a collaboration between astronomers at the University of 
Arizona, the University of California at San Diego, New York University, and 
Princeton University.  A companion paper (Cool et al. in preparation) 
describes the 
data reduction, redshift fitting algorithm, redshift assessment and 
completeness, as well as lessons learned from carrying out the survey.
The outline of this paper is as follows.  Section~\ref{sec:motivation} describes
the motivation for and goals of our survey.  Section~\ref{sec:prism} summarizes 
the hardware built for and used by the survey, namely the prism and IMACS 
camera.  Section~\ref{sec:fields} provides information about the fields 
observed by the survey, while Section~\ref{sec:target_selection} outlines the 
algorithms used to select targets for slitmasks and 
Section~\ref{sec:mask_design} presents the mask design considerations. 
Section~\ref{sec:observations} describes how the observations were carried 
out, and Section~\ref{sec:photomerge} provides 
detailed information on the various optical 
photometric datasets used in each field, 
including derived zeropoints. A summary of the observed PRIMUS 
data and a comparison to other surveys 
is provided in Section~\ref{sec:data_summary}.  
Throughout the paper, AB magnitudes are used.

\section{The Motivation for Low Resolution Spectroscopy} \label{sec:motivation}

High-precision spectroscopic redshifts are the gold standard for
determining distances to galaxies outside the local Universe. 
However, conventional spectroscopy of large numbers of faint galaxies is
observationally extremely expensive, particularly over the wide areas
needed to suppress the effects of cosmic variance --- the fact that
large-scale structure can greatly affect the density in any
particular, small patch of the Universe we map.  Existing deep,
wide-field, multi-wavelength imaging exceeds its spectroscopic reach by
at least a factor of 10, and conducting a survey of hundreds of thousands 
of galaxies
to $i=23$ would require a mammoth and unfeasible investment of time
for a high resolution spectroscopic survey on the current, existing
instruments. Low-precision broad-band photometric redshifts 
are commonly used to fill this gap and have become popular because they
are much easier to acquire and therefore can be done for much larger
data sets \citep[e.g.,][]{Budavari03,Mobasher04, Padmanabhan05,
Gladders05, Pell09,Adami10}. However, the usual broad-band
($\mathcal{R}\equiv\lambda/\Delta\lambda\sim3$--$5$) photometric redshifts are
susceptible to systematic errors, have a non-trivial fraction of
catastrophic failures ($>$10\%), and typically have $\sim$5\% redshift errors, 
which are large compared to the physical clustering lengths. 
Several recent efforts involve combining both optical and NIR imaging,
though the redshift error is still $\sim$4-5\% 
\citep{Brodwin06,Rowan-Robinson08}.
The best photometric redshifts published that use several 
broad-band filters are those from CFHTLS, which have $\sim$3-4\%
redshift errors and $\sim$2-4\% catastrophic outliers (defined as
$\ddzz>0.15$), depending 
on the depth of the sample \citep{Coupon09}; these rates are 
better than other surveys due to the depth of the imaging data.

Systematic errors on the redshifts are often the limiting uncertainty 
in the quantification of galaxy properties and clustering from surveys 
that rely on this technique.  
Errors of a few percent in distance produce significant uncertainties in 
flux-dependent quantities such as stellar mass and SFR, and they create
punishing background subtraction problems for clustering and environment
analyses of any sort.  Errors of 0.5\% alleviate all of these problems, 
especially when studying clustering statistics. For example, at $z=0.5$ an 
error of 5\% in $\ddzz$ leads to an error of 0.4 magnitudes in the distance 
modulus, while an error of 0.5\% in $\ddzz$ leads to an error of 0.04 
magnitudes.

The COMBO-17 survey \citep{Wolf03,Wolf04} achieved considerable
success with a hybrid technique, in which twelve intermediate-band
optical images plus five broad-band images were used to create a
$\mathcal{R}\approx$10-15 spectrum.  Having somewhat higher resolution allows
COMBO-17 to identify the spectral energy distribution (SED) breaks
more accurately, thereby improving their redshift precision to
$\sim$3\% and reducing the number of outliers.  The COSMOS team has
since followed suit and obtained 30 bands of photometry over 2 deg$^2$ 
to obtain $\sim$0.7\% redshifts to $z\sim1$ at a depth of 
$i=22.5$ \citep{Ilbert09}.  For reference, the imaging for 20 of 
these photometric bands was obtained at the
Subaru 8m telescope and required 20 nights of observing.

However, these surveys with more than ten bands of photometry 
cover only 1--2 deg$^2$ of sky, as the
exposure time of a intermedian-band imaging survey scales roughly as $\mathcal{R}^2$:
one factor of $\mathcal{R}$ for the number of filters, the other for the length
of each exposure to reach a given signal-to-noise (S/N) ratio. This
scaling means that even wide-field photometric surveys are highly
inefficient, and the current best photometric redshift surveys over wide areas 
have a precision of 3-4\% \citep[CFHTLS, ][]{Coupon09}.
 The major source of inefficiency is that nearly all of
the pixels to $i_{\rm AB}=23$ 
are blank sky.  It is much more efficient to disperse the
light and use more of the detector pixels on sources.  
With dispersion, exposure
times scale only as one power of $\mathcal{R}$ to reach a given S/N per
resolution element.  For $\mathcal{R}\approx 15$, this factor is more than
sufficient to pay off the throughput penalties of the spectrograph or
the need to use two masks to mitigate slit collisions.

With PRIMUS, we have used the IMACS instrument on the 6.5m Magellan
telescope to obtain both a significantly higher survey speed and
spectroscopic resolution ($\mathcal{R} \approx 40$) than even the best
photometric redshift surveys, e.g., COMBO-17 and the COSMOS 30-band 
photometric redshifts.  By using the detector
pixels to multiplex spectrally instead of integrating on blank sky, we
cut the survey time by a substantial factor while gaining higher
spectral resolution.  

%We are therefore more efficient and obtain more
%precise redshifts than photometric redshift surveys.  
In general, the advantage of low-resolution spectroscopy compared
to intermediate-band photometric redshifts depends on the relative
fields of view, throughputs, and apertures of the facilities as
well as on the target density and completeness requirements for the
science goals. Higher target densities favor imaging surveys, because
there is less blank sky; lower target densities favor spectroscopic
surveys, as one can fill more of the sky with higher spectral resolution
data. Some science cases depend sensitively on high survey completeness;
others require only statistical control of the incompleteness.
Additional less important factors are that, compared to spectroscopy,
photometry allows one to optimally weight the radial profile of the
target and to avoid losing light at a slit. Conversely, spectroscopy
allows one to obtain continuous spectral sampling, as opposed to
fixed filters, and to optimally weight the spectral information,
e.g., to downweight information at the wavelength of sky lines.

As we show in
Section~\ref{sec:data_summary} and in Cool et al. (in preparation), 
our spectra result in redshifts precise to $\dzz=0.005$. 
They also provide medium-band SEDs and detection of the strongest
emission lines and broad AGN lines.  Given the relatively low dispersion 
of our spectra, we do not attempt to measure detailed line
diagnostics for individual sources 
nor the internal velocity distributions of groups and
clusters, given our redshift precision.  
However, redshift-space distortions smear out the cosmic
web by $\sim500\,\mathrm{km\,s^{-1}}$, such that redshifts more
accurate than about 1\% do not improve most clustering or environment
analyses, which integrate over these distances anyway.  For our
science goals, therefore, exchanging higher resolution redshifts for a
faster, wider redshift machine is a very worthwhile trade, and is the
only currently feasible way to obtain redshifts of large samples over 
wide areas of sky.  Future very wide-field spectroscopic instruments 
may provide an 
alternative route, though these spectrographs do not currently exist.
We emphasize that any existing wide-field multi-object spectrograph would
benefit from multiplexing using a low-dispersion prism.  For PRIMUS,
IMACS on Magellan is ideally suited given the relatively large 
field of view of 0.18 deg$^2$.

We use the near-IR, mid-IR, UV, and X-ray data in our fields as proxies for 
stellar mass, star formation, and AGN activity.  Optical emission and 
absorption lines provide alternative approaches to these measurements, but 
require spectra of far better quality than what is needed to measure a 
redshift.  Given the existence of the multi-wavelength imaging in our 
fields, our goal is to efficiently obtain spectra over a wide area 
in order to determine redshift measurements.

\section{Prism and Camera Characteristics} \label{sec:prism}

\begin{figure}[t]
\plotone{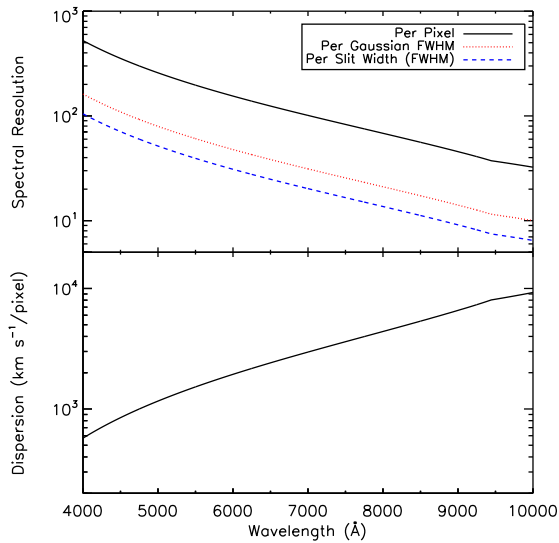}
\caption{\label{fig:resolution}
\small 
Top: Resolution $\lambda/\Delta\lambda$ for the prism, quoted per
pixel, per Gaussian FWHM (the standard deviation of the dispersion function 
divided by 2.35), and per 5-pixel (1\arcsec) slit.  
Bottom: The dispersion in km ${\rm s^{-1}}$ per pixel.
Both figures are for the performance in the center of the field of view;
there are slight changes toward the edges, negligible for performance.
}
\end{figure}

\begin{table*}[t]
\tablewidth{0pt}
\begin{center}
\label{tab:fieldstab}
\small
\caption{PRIMUS coverage statistics for science fields.}
\begin{tabular}{lcccrrrrrrccc}
\cr
\colrule
\colrule
\vspace{-3 mm} \cr
Field name & RA\footnotemark[a] & Dec & Area\footnotemark[b]  & $N_{\rm{mask}}$\footnotemark[c] & 
$N_{\mathrm{total}}$\footnotemark[d] & $N_{\mathrm{primary}}$ & $N_{\mathrm{stars}}$\footnotemark[e] &
$N_{\mathrm{gal+AGN}}$ &
$N_{\mathrm{primary}}$& {\it Spitzer}\footnotemark[f] & {\it GALEX} & X-ray \cr
& (J2000) & (J2000) & (deg$^2$) && \multicolumn{2}{c}{observed} & \multicolumn{3}{c}{robust z} &
 \multicolumn{3}{c}{(deg$^2$)} \cr
\vspace{-3 mm} \cr
\colrule 
\colrule 
\vspace{-3 mm} \cr
CDFS-SWIRE \ &  03:32 & -28:54 & 1.95 & 29 \ &  46,127 \ & 33,970 \ &   933  & 22,898 & 19,942  & 1.94 & 1.95 &      \cr
COSMOS       &  10:00 & +02:21 & 1.03 & 18 \ &  39,971 \ & 20,741 \ & 2,162  & 17,963 & 11,230  & 1.03 & 1.03 & 1.03 \cr
DEEP2 02hr   &  02:30 & +00:36 & 0.58 &  8 \ &  14,213 \ &  9,814 \ &    99  &  7,289 &  5,778  & 0.47 & 0.31 & 0.52 \cr
DEEP2 23hr   &  23:30 & +00:09 & 0.67 & 10 \ &  15,604 \ & 10,491 \ &    87  &  7,329 &  5,811  &      & 0.67 & 0.67 \cr
DLS F5       &  13:58 & -11:18 & 1.04 & 13 \ &  30,653 \ & 22,895 \ & 1,218  & 11,032 & 10,147  &      &      &      \cr
ELAIS S1     &  00:36 & -43:30 & 0.90 & 12 \ &  21,534 \ & 15,839 \ &   842  &  9,992 &  9,101  & 0.90 & 0.86 & 0.52 \cr
XMM-LSS      &  02:20 & -04:45 & 2.88 & 42 \ & 102,218 \ & 61,689 \ & 3,364  & 44,451 & 34,590  & 2.84 & 2.65 & 2.77 \cr 
\vspace{-3 mm} \cr
\colrule                      
\vspace{-3 mm} \cr                                          
Total        &        &       & 9.05 & 132 \ & 270,320 \ & 175,439 \ & 8,705 & 120,954 & 96,599 & 7.18 & 7.47 & 5.51
\end{tabular}
\footnotetext[a]{Approximate center of field observed by PRIMUS}
\footnotetext[b]{Area over which PRIMUS obtained spectroscopy; regions not surveyed due to bright stars, 
missing photometry, and CCD chip gaps have been removed}
\footnotetext[c]{Number of slitmasks observed}
\footnotetext[d]{Number of primary and total targets observed, duplicate observations are not included 
(see text for details)}
\footnotetext[e]{Number of stars, galaxies+AGN, and primary objects (stars+galaxies+AGN) with robust redshifts, defined as Q=3 or 4 (see Section 9)}
\footnotetext[f]{Area with joint PRIMUS and either {\it Spitzer, GALEX, Chandra}  or {\it XMM} coverage.}
\end{center}
\end{table*}

PRIMUS was conducted with the
Inamori Magellan Areal Camera and Spectrograph \citep[IMACS,][]{Bigelow03}
operating on the Magellan I (Baade) Telescope at Las Campanas Observatory.
The IMACS f/2 camera delivers a $27.2\arcmin$ square field of view with slight
vignetting, for an active area of 0.18 deg$^2$.  
The pixel scale is $0.2\arcsec$ per pixel, so our spectra are well-sampled.
The detector has eight 2k$\times$4k$\times$15 micron CCDs.  The chip gaps
are 11.4$\arcsec$ between the short sides of the CCDs and 18.4$\arcsec$ 
between the long
sides.  Using the fast readout mode, the gain is $\sim0.8$ e$^-$/adu and 
the read noise is $\sim$4 e$^-$/pixel. 

As with traditional spectroscopy programs, PRIMUS used custom-designed
slitmasks, made with the laser mask miller at Las Campanas.  The bulk of
the survey used a slit size of $1.0\arcsec$ wide by $1.6\arcsec$ long.
Details of the mask design are given in Section \ref{sec:mask_design}.
The innovative design aspect of PRIMUS is to use a prism in the place of 
the transmission gratings normally used with the f/2 camera.
The prism was designed and fabricated for our survey; it was mounted in IMACS
in January 2006 and was immediately made available to general users.
The prism is made of three pieces of glass,
one piece of PBM2Y glass in between two pieces of FSL5Y glass 
(all i-line glasses from O'Hara).  The size is 190 mm in diameter by 84mm 
depth at its thickest point.
The total cost of the prism and mounting was $\sim\$20$K.

As a dispersing element, a prism is ideal in that it has a
very broad transmission curve, extremely high throughput, 
requires no order blocking filters,
and does not exhibit excessive scattered light compared to 
transmission gratings.  Additionally, because of the wide field of 
view of the f/2
camera, grisms operating at 100 lines mm$^{-1}$ and below will be 
strongly affected by zero-order direct images of the slits.
A prism does not suffer from this effect. It also has no blaze condition and
has a higher throughput over a wider wavelength range compared to gratings
and grisms.
The low-dispersion prism designed for PRIMUS 
yields spectra from 3900\AA\ to 1 $\mu$m that requires only
150 pixels in the spectral direction, a tiny fraction (2\%) of the 
IMACS f/2 field of view.  We are therefore able to multiplex both spatially
and spectrally to observe $\sim2500$ galaxies at once.  
Figure \ref{fig:resolution} shows the wavelength-dependent 
spectral resolution (top panel) and dispersion (bottom panel) 
of the prism, which changes considerably from blue to red.  
The throughput of the prism is $\sim90$\%, and 
the overall throughput of the prism plus camera, including the telescope but 
not the atmosphere, peaks at 24\% at 6900 \AA, is 20\% 
 at 5200 and 7600 \AA, and is 10\% at 4450 and 8500 \AA.

\section{Fields} \label{sec:fields}

\begin{figure*}[t]
\plotone{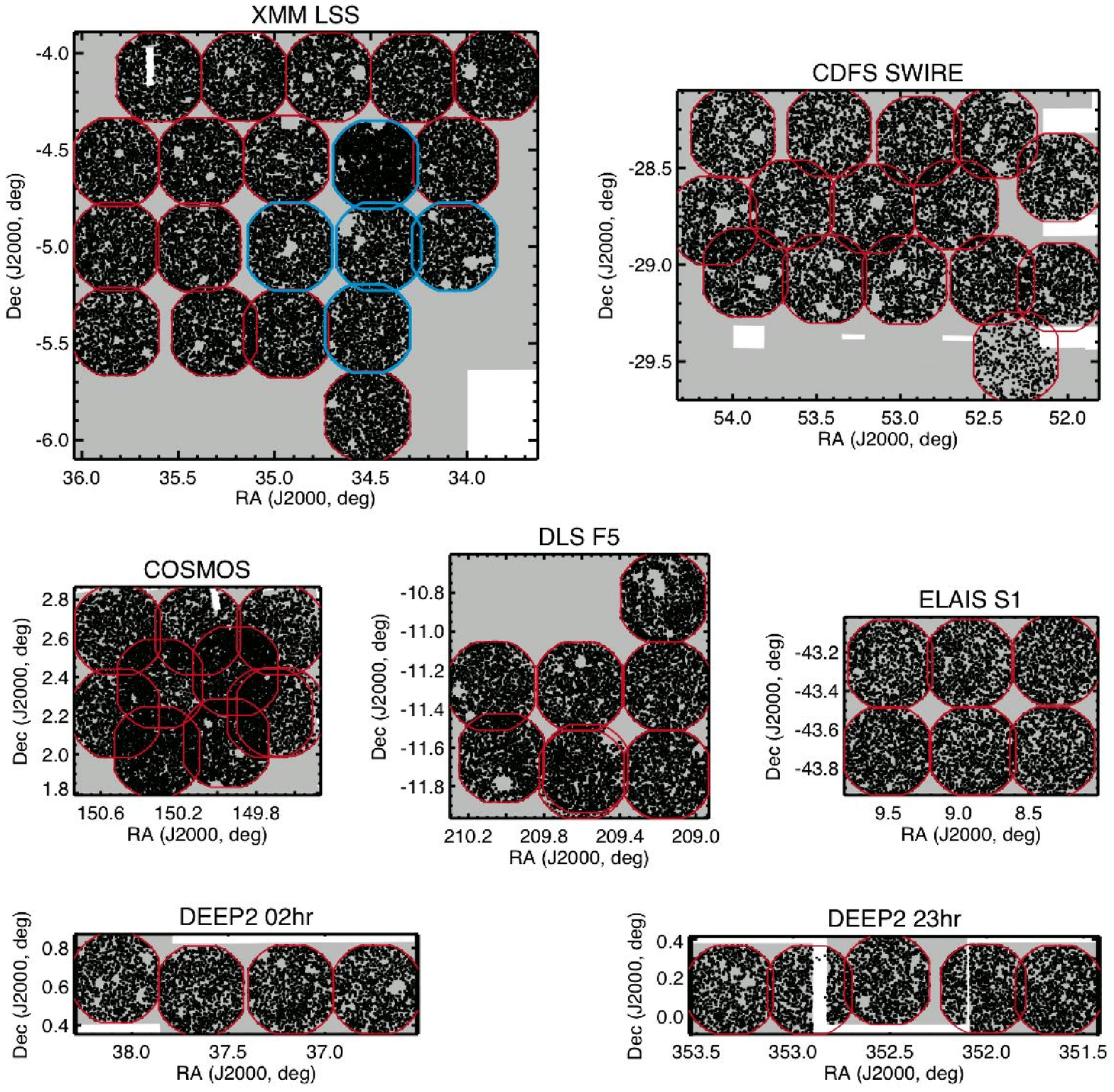}
\caption{\label{fig:coverage}
\small PRIMUS coverage of the various science fields targeted.  Grey
shading indicates the area surveyed by photometric catalogs used for
targeting, black crosses are PRIMUS targets, and red lines show the
outline of the observed PRIMUS slitmasks in the field.  The area
outlined in cyan in the XMM-LSS field is the SXDS region; see text for
details.  Only 20\% of the PRIMUS targets are shown here for clarity;
the actual sampling density is five times higher than shown.}
\end{figure*}

\begin{figure}[t]
\plotone{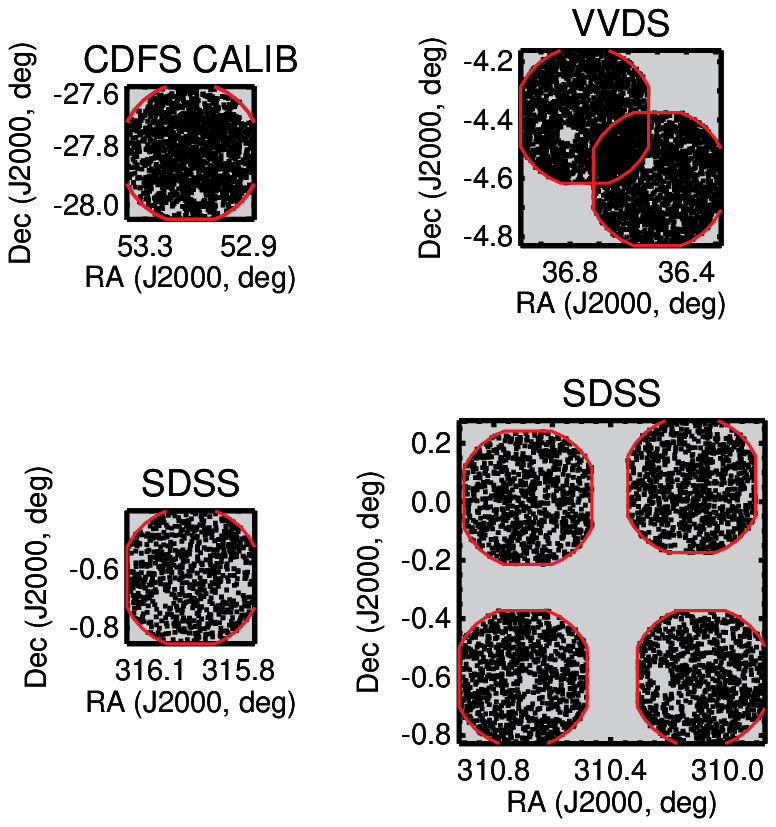}
\caption{\label{fig:coverage_calib}
\small PRIMUS coverage of the various calibration fields targeted.  
Grey shading indicates the 
area covered by photometric catalogs used for targeting, 
black crosses are PRIMUS targets, and red lines show the outline 
of the observed PRIMUS slitmasks in the field.  Only 50\% of the PRIMUS
targets are shown here for clarity; the actual sampling density in these
calibration fields is two times higher than shown.
}
\end{figure}

PRIMUS primarily targeted regions of the sky with existing deep multi-wavelength 
imaging.  Our goal was to cover all of the southern SWIRE fields that had 
pre-existing optical photometry from which we could design slitmasks.  In order 
to facilitate being able to observe for full nights over roughly half of the year, 
we also observed several non-SWIRE fields, some of which have deep {\it Spitzer} 
data.  All but one of our fields have {\it GALEX} coverage and all but two have 
X-ray coverage from either {\it Chandra} or {\it XMM}.  These fields are
designated as ``science fields.''  We also targeted a 
small number of ``calibration fields'' that have existing 
high-resolution spectroscopic redshifts from other surveys, which we 
use to characterize our redshift accuracy and precision.

PRIMUS observed a total of ten independent fields, seven science 
and three calibration fields.  
The science fields include the Chandra Deep Field South-SWIRE
field \citep[CDFS-SWIRE, which is not centered on the CDFS proper,][]{Giacconi01}, the 23$^{hr}$ 
and 02$^{hr}$ DEEP2 fields, the COSMOS field, the European Large Area ISO Survey -
South 1 field \citep[ELAIS-S1,][]{Oliver00}, the XMM-Large Scale Structure 
Survey field \citep[XMM-LSS,][]{Pierre04}, and
the Deep Lens Survey \citep[DLS,][]{Wittman02} F5 field.  Of these,
SWIRE coverage exists for the CDFS, ELAIS-S1, and XMM-LSS fields, and
S-COSMOS \citep{Sanders07} 
provides deep {\it Spitzer} coverage of the COSMOS field.
X-ray coverage exists for both DEEP2 fields ({\it Chandra}), the XMM-LSS
field ({\it XMM}), and the COSMOS and ELAIS-S1 fields (both {\it Chandra} and 
{\it XMM}).  The total area with joint PRIMUS and either {\it Spitzer, GALEX} 
or X-ray coverage is 7.18 deg$^2$, 7.47 deg$^2$, and 5.51 deg$^2$, respectively.

Details of the PRIMUS science fields are given in Table 1.
Listed are the approximate field centers as observed by PRIMUS, the total
area targeted, the number of slitmasks observed, and the number of unique
targets and robust redshifts derived in each field, along with the area of
each field that contains multi-wavelength coverage.  We separately list 
the number of stars versus galaxies and AGN that have robust redshifts.
Robust redshifts are defined as having a redshift confidence flag Q$=$3 or
4; see Section 9 for details.
The numbers listed for the primary sample include stars, galaxies, and AGN.
The areas listed in Table 1 are the total area
that has PRIMUS targets and are conservative in that they do not
include regions masked by bright stars, missing photometry, 
or that fell in between chip gaps on the CCD.  If these areas are included
then the total area of the survey is 10.78 deg$^2$.

Table 2 provides information on the PRIMUS coverage of the calibration fields, 
which include the original CDFS (which has high-resolution 
spectroscopic redshifts from the VVDS), the deep VVDS field, and a portion of 
the SDSS equatorial Stripe 82 \citep{Aba09}. 
The DEEP2 02$^{hr}$ and 23$^{hr}$ fields and the COSMOS field are used as 
both science and calibration fields.

Figures \ref{fig:coverage} and \ref{fig:coverage_calib} show the PRIMUS sky coverage
in our science and calibration fields.  The light grey areas indicate photometric 
coverage of the catalog used for targeting, red lines 
outline the observed PRIMUS slitmasks, and the black crosses show PRIMUS 
targets, where we plot a random 20\% of the total targets for clarity in the 
science fields and 50\% of the targets in the calibration fields.  
In the XMM-LSS field, five PRIMUS pointings are outlined in cyan; in this
area we used photometry from the Subaru/XMM-Newton Deep Survey \citep[SXDS,][]{Furusawa08}
 for targeting (see the next section for 
details). During targeting, areas around bright stars were masked to avoid 
photometric errors. These regions can be seen as grey areas within the PRIMUS 
masks that do not have targets.

\begin{table*}[t]
\tablewidth{0pt}
\begin{center}
\label{tab:calibfieldstab}
\small
\caption{PRIMUS coverage statistics for calibration fields.}
\begin{tabular}{lcccrrrrrr}
\cr
\colrule
\colrule
\vspace{-3 mm} \cr
Field name & RA\footnotemark[a] & Dec & Area & $N_{\rm{mask}}$ & 
$N_{\mathrm{total}}$ & $N_{\mathrm{primary}}$ & $N_{\mathrm{stars}}$ &
$N_{\mathrm{gal+AGN}}$ &  
$N_{\mathrm{primary}}$ \cr
& (J2000) & (J2000) & (deg$^2$) && \multicolumn{2}{c}{observed} & \multicolumn{3}{c}{robust z} \cr
\vspace{-3 mm} \cr
\colrule 
\colrule 
\vspace{-3 mm} \cr
CDFS-CALIB & 03:32 & -27:49 &  0.15 & 3 \ \ & 1,841  & 0     \ &    82 & 1,197  & 0       \cr
SDSS       & 20:52 & -00:18 &  0.74 & 5 \ \ & 7,571  & 6,100 \ & 1,773 & 2,572  & 3,534   \cr
VVDS       & 02:26 & -04:36 &  0.27 & 2 \ \ & 5,876  & 0     \ &   132 & 2,672  & 0       \cr
\vspace{-3 mm} \cr
\colrule
\vspace{-3 mm} \cr
Total      &       &        &  1.16 & 10 \ \ & 15,288 & 6,100 \ & 1,987 & 6,441 &  3,534
\end{tabular}
\footnotetext[a]{Column meanings are identical to Table 1}
\end{center}
\end{table*}

\begin{table*}[t]
\label{tab:target_phot}
\begin{center}
\small
\caption{\small Targeting photometry and depth of primary sample in each field.}
\begin{tabular}{llcccl}
\cr
\colrule
\vspace{-3 mm} \cr
Field name     & Telescope   & 100\% sampling\footnotemark[a] & 30\% sampling  & Non-primary limiting & reference\footnotemark[b] \cr
               &             &  mag (AB)                      & mag range (AB) & mag (AB) & \cr
\vspace{-3 mm} \cr
\colrule 
\colrule 
\vspace{-3 mm} \cr
CDFS-SWIRE\footnotemark[c]     & CTIO/Blanco & $i<22.5$  & $22.5<i<23.0$  & $i<23.5$ & \cite{Lonsdale03} \cr
COSMOS         & HST/ACS     & $I$\footnotemark[d]$<22.7$ & $22.7<I<23.2$  & $I<23.2$ & \cite{Koekemoer07} \cr
DEEP2 02hr     & CFHT        & $R<22.8$  & $22.8<R<23.3$  & $R<24.1$ & \cite{Coil04} \cr
DEEP2 23hr     & CFHT        & $R<22.8$  & $22.8<R<23.3$  & $R<24.1$ & \cite{Coil04} \cr
DLS F5         & CTIO/Blanco & $R<22.8$  & $22.8<R<23.3$  & $R<23.3$ & \cite{Wittman06} \cr
ELAIS S1\footnotemark[c]       & ESO/WFI     & $R<22.7$  & $22.7<R<23.2$  & $R<23.7$ & \cite{Berta06} \cr
XMM-LSS/CFHTLS & CFHT        & $i<22.5$  & $22.5<i<23.0$  & $i<23.5$ & T0003 release \footnotemark[e] \cr
XMM-LSS/SXDS   & Subaru      & $i'<22.5$ & $22.5<i'<23.0$ & $i'<23.5$ & \cite{Furusawa08} \cr
\vspace{-3 mm} \cr
\colrule
\end{tabular}
\footnotetext[a]{Photometric band and depth of primary sample used for spectroscopic target selection.  See \S5.1.1 for details.}
\footnotetext[b]{Reference for photometric catalog}
\footnotetext[c]{These fields have an additional IRAC 3.6 \micron \ flux limit of $\sim$10 $\micron$Jy.  See \S5.1 for details.}
\footnotetext[d]{F814W band}
\footnotetext[e]{http://terapix.iap.fr}
\end{center}
\end{table*}

\section{Target Selection} \label{sec:target_selection}

\subsection{Overview of Target Selection}

The goal of PRIMUS was to obtain redshifts for a flux-limited, high-density 
sample of galaxies, spanning all galaxy types, over a wide area to
$z\sim1.0$.  Pushing to $z>1$ would require higher-resolution
spectroscopy than was available with the low-dispersion prism that we
used.  Additionally, one needs to sample redward of the 4000 \AA \ break 
reasonably well, and the spectral resolution is particularly low at the red end
of our spectra (see Figure~\ref{fig:resolution}).

We used existing photometry in each field to choose targets for
spectroscopy.  The imaging data used in each field for targeting 
is at least one magnitude deeper than our targeting limit.
As the imaging was taken from several different sources
(using different cameras on different telescopes), it is not expected to be 
perfectly homogeneous.  The differences in targeting photometric band 
and depth can and will be accounted for when correcting for the survey 
completeness and will be performed for all relevant science papers.

In order to maximize the area of the survey, 
thereby minimizing cosmic variance errors, we limited our total
exposure time per slitmask to 1 hour.  In the COSMOS field we
exposed for 1.5 hours to reach a higher S/N in our spectra and 
to target to a slightly fainter depth, due to the high-quality, deep 
ancillary multi-wavelength data and HST imaging in that field.  

We generally targeted all galaxies to $i<22.5$ and sparse-sampled 
$22.5<i<23$ objects with well-defined {\it a priori} sampling rates.  
In this way PRIMUS would not be dominated by galaxies at our faintest 
flux limits.  Because the targeting and selection weights were all tracked and 
saved, a statistical complete sample to $i<23$ can be 
created from the PRIMUS data.  
As the low-dispersion spectroscopy technique we used was experimental,
we intentionally targetted fainter galaxies, with the intention of 
determining the redshift success thresholds from the spectroscopic
data.  The result is that our success rate is a function of magnitude 
(as discussed below and in detail in Cool et al. in preparation,
and we failed on a fair fraction of the fainter targets below $i=22.5$.

Three fields have SWIRE coverage: CDSF-SWIRE, ELAIS-S1, and XMM-LSS.
In the CDFS-SWIRE and ELAIS-S1 fields (but not the
XMM-LSS field), due to a bug in the design code, at $i>23.0$ only objects with
IRAC counterparts brighter than 10 $\micron$Jy at 3.6 \micron \ in the
SWIRE catalog were targeted.  This leads to an additional selection
effect that will be accounted for in relevant science papers.  
In all three of the SWIRE fields, we intentionally 
additionally targeted sources to $i<23.0$ that were bright at 3.6
\micron \ ($>$10 $\micron$Jy in a 3.8$\arcsec$ aperture), and sources
to $i<23.5$ that were detected in the IRAC 5.8, 8.0, or MIPS 24
\micron \ bands.  
In the XMM-LSS field, we initially designed slitmasks in the SXDS area of the
field, which had deep publicly-available Subaru imaging prior to the
first CFHTLS imaging release; these masks are outlined in cyan in
Figure 2.  In the rest of the XMM-LSS field, we used CFHTLS imaging
for targeting.

In fields that lacked available $i$ or $I$ band imaging when we
designed our slitmasks, we used $R$ band photometry and adjusted the 
magnitude selection limit such that cumulative galaxy number counts 
at the magnitude limit were comparable across fields.  This procedure 
corrects for the average $R-i$ color difference of galaxies.  
We used the star-galaxy separation that was available in the
targeting catalogs in each field; the details of this separation are
therefore field-dependent.

\subsubsection{Primary Sample}

Table 3 lists information on the photometry
that was used in each field to select targets for spectroscopy and the 
limiting apparent magnitude of the primary sample.
The ``primary'' sample in each field is defined as those objects from
which we can create a statistically complete sample.  These objects
have a well-understood selection function both in terms of their selection
magnitude and their location in the plane of the sky, i.e. they have a
recoverable spatial selection.  This sample can therefore be used for
statistical measurements such as the two-point correlation function and
luminosity function, as there is little to no bias in terms of the
spatial density in this sample.  

\subsubsection{Sparse Sampling}

From the input photometric catalog used for targeting, we defined two selection
weights for each object: (1) a magnitude-dependent sparse-sampling weight, 
which was used to select a fraction of galaxies at random in the 0.5 
magnitude interval above our primary sample targeting limit; 
and (2) a spatial density-dependent weight, which was used to select 
galaxies in crowded regions of the sky where the spectra of adjacent galaxies
would overlap on the detector.  We refer to these two selection
weights as the (magnitude-dependent) ``sparse-sampling weight'' and
the (spatially-dependent) ``density-dependent weight.''  Details of
the density-dependent sampling are given in Section \ref{sec:density}.

The sparse-sampling weight is equal to 1 to the ``100\% sampling 
limiting magnitude'' 
in each field (generally $i=22.5$) and equal to either 1 (if spatially 
isolated) or 0.3 (if there are collisions) in the 
``30\% sampling magnitude range'' (generally 0.5 magnitude fainter).  In 
defining these weights, we first checked to see if an object had two 
nearby neighbors that would have
created a collision in designing the slitmasks (see Section \ref{sec:density}
for details on the definition of nearby neighbors).  If an object did not 
have collisions, then it was not sparse-sampled (i.e. the sparse-sampling
weight was 1).  If an object had two or more nearby neighbors, 
it was sparse-sampled with a weight of 0.3.

\subsubsection{AGN Selection}

When choosing galaxies, we kept point sources if their optical and/or IR 
colors indicated that they may be AGN.  Specifically, point 
sources that passed any of the following {\it Spitzer} color cuts were targeted:
    \begin{eqnarray}
    &&[3.6]-[4.5] > -0.1 {\rm \ or} \nonumber \\ 
    &&[3.6]-[5.8] > -0.4 {\rm \ or}  \nonumber \\ 
    && [3.6]-[8.0] > -1.0  {\rm \ or} \nonumber \\ 
    &&[3.6]-[24] \ > -2.0 
\label{eqn:swirestar1}
    \end{eqnarray}

Additional optical color thresholds were used to define AGN candidates in 
most fields, where the color definitions used depended on the available imaging
in each passband.
In the CDFS, point sources that passed either of the following optical 
cuts were targeted:
\begin{eqnarray}
&& g-r<0.9 {\rm \ and \ } \nonumber \\
&& \hspace{10mm} |(r-i)-0.45(g-r)-0.03|>0.2 \\
&& u-g<0.7 {\rm \ and \ } g-r<1.0
\end{eqnarray}

In the COSMOS field, point sources in the HST imaging with $19<I<23.2$ 
were targeted, with no color pre-selection applied.
In the two DEEP2 fields, we targeted photometrically-selected
 AGN with $g<21$ and $g-i<2$ from \cite{Richards04}.
In the SXDS area of the XMM-LSS field, 
objects that were detected by XMM were selected to $i<23.5$. 
In addition, point sources were targeted if their $VRI$ optical colors indicated
that they may be AGN, where the following selection was used:
\begin{equation}
 V-R<0.9 {\rm \ and \ } |(R-I)-0.45(V-R)-0.03|>0.2
\end{equation}
In the rest of the XMM-LSS field, point sources with $17.5<i<21.0$ 
that passed either of the following optical 
cuts were targeted:
\begin{eqnarray}
&& g-r<1.0 {\rm \ and \ } \nonumber \\
&& \hspace{15mm} |(r-i)-0.5 (g-r)|>0.1 \\
&& u-g<0.6
\end{eqnarray}

While the targeting weights of AGN identified by their
optical colors vary from field to field, for science purposes 
we will be able to perform completeness corrections within each field.  
In general, the X-ray imaging data that exists in these fields will 
be used to define parent AGN samples.  We can then account for our 
targeting selection effects to define a complete sample of X-ray 
detected AGN.

In instances where the team that provided the 
targeting photometry requested spectroscopy of small numbers of
 specific objects, they were added as high priority filler targets.

\subsubsection{Additional Targets}

We additionally targeted galaxies at lower priority that did not pass the
density-dependent sampling or magnitude-dependent
sparse-sampling, often to 0.5 magnitude fainter than the ``30\% sampling 
magnitude limit'', if they could be placed on slitmasks without 
colliding with higher priority, primary targets.
Empty slits with no objects were milled in all fields
except the DEEP2 02hr and COSMOS fields, to test sky subtraction and
extraction. We also observed photometrically-selected F stars as
spectrophotometric calibrators on each slitmask,
where SDSS imaging was used to select F stars where available.
See Cool et al. (in preparation) for information on the data reduction and 
flux calibration of our spectra.

\begin{figure*}[t]
\plotone{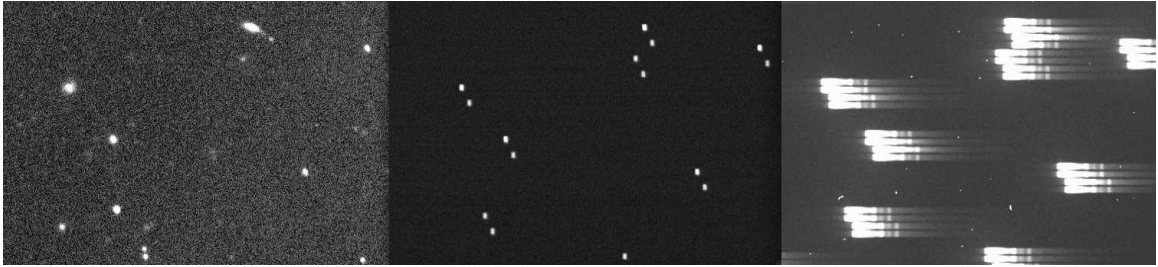}
\caption{\label{fig:zoom}
\small Small portion (roughly 1.25\arcmin \ wide) 
of imaging in one PRIMUS field (left), the corresponding 
slit assignment for one of the two slitmasks that covers this region (middle), 
and the raw PRIMUS data from a single exposure of this slitmask (right), with 
wavelength increasing to the left. 
Note that the image in the middle panel is not taken on sky; it is a
non-dispersed image of a portion of the slitmask taken with a calibration 
lamp. 
Using nod-and-shuffle,
each object has four traces on the CCD, as seen in the right panel; 
two contain the object and two contain shuffled sky spectra.
}
\end{figure*}

\subsection{Selection Details in Calibration Fields}
\label{sec:sel_details}

Target selection in our calibration fields depended on the 
available high-resolution spectroscopic data.  Details of the 
target selection for each of our calibration fields is given 
below.

\subsubsection{CDFS-CALIB}
Three slitmasks were observed in the original CDFS, which has high-resolution
spectroscopic redshifts from the VVDS.  Priority was given to targets
with high-quality VVDS redshifts \citep[rank 3 or higher,][]{Lefevre05} 
to $i=23.3$.  We also
targeted galaxies with lower-quality VVDS redshifts (rank 1 or 2), 
as well as objects with spectroscopic redshifts from the Intermediate Mass 
Galaxy Evolution Sequence survey \citep[IMAGES, ][]{Ravikumar07} (rank 1 or 2), 
and COMBO-17 galaxies with $20<R<21.5$.

\subsubsection{DEEP2}
The two fall DEEP2 fields are used by PRIMUS for both science and calibration.
In these fields, high-resolution spectroscopy is available from the 
DEEP2 survey for galaxies selected in $BRI$ color-color space to be at
$z>0.7$ to a depth of $R=24.1$ \footnote{http://deep.berkeley.edu/DR3}. 

In the region of the 02hr field that has DEEP2 spectroscopy 
(roughly 75\% of the photometric area available), 
PRIMUS targeted a set of galaxies that is complementary to the DEEP2 
sample.  PRIMUS targeted galaxies with $BRI$ colors indicating 
redshifts $z<0.7$ (i.e. exactly those not
targeted by DEEP2) with no sparse-sampling to $R<22.8$ and 100\% of
the non-collided and 30\% of the collided galaxies with $22.8<R<23.3$.
Slitmasks were then filled with galaxies that had DEEP2 
redshifts to $R<24.1$, to use in characterizing our redshift precision,  
as well as the rest of the $z<0.7$ galaxies to $R<24.1$ that were 
not initially selected by the density-dependent sampling.

In the region of the 02hr field that did {\it not} have
DEEP2 spectroscopy (roughly 25\% of the PRIMUS area), all galaxies
were targeted to $R<22.8$ and 100\% of non-collided and 30\% of collided
galaxies with $22.8<R<23.3$, without any $BRI$ color pre-selection.
Slitmasks were filled with galaxies to $R<24.1$ not initially selected by 
the density-dependent sampling.  

In the DEEP2 23hr field in 40\% of the area covered by PRIMUS we targeted all 
$BRI$ color pre-selected galaxies at $z<0.7$ to $R<23$ and 30\% of
all galaxies with $23<R<23.5$.  In the other 60\% of the PRIMUS area 
we targeted 0.2 magnitudes brighter: 100\% of galaxies to $R<22.8$ and 
100\% of non-collided and 30\% of collided galaxies with $22.8<R<23.3$.  
Slitmasks were filled with galaxies
with DEEP2 redshifts to $R<24.1$ and the rest of the $z<0.7$ 
galaxies not initially selected by the density-dependent sampling.

In the DEEP2 fields one can therefore 
coadd the primary PRIMUS sample and the DEEP2 
sample, using the normal survey selection functions, to create a full galaxy
sample to $R=23.3$.

\subsubsection{SDSS}
In the SDSS calibration field we observed one PRIMUS slitmask per pointing, 
instead of two slitmasks per pointing as in the other fields.  We targeted
galaxies (at 100\% sparse-sampling) and stars fainter than $r=21$ 
(at 50\% sparse-sampling) to $r<22$ or $i<21$.  Targeting weights were
saved for this primary sample.  We additionally targeted the remaining 
faint stars and galaxies that did not pass the 
density-dependent sampling; these sources are not in the primary sample.
 As in the DEEP2 fields, we 
targeted photometrically-selected AGN with $g<21$ and $i>19$ from 
\cite{Richards04}.

\subsubsection{VVDS}
Two calibration slitmasks were observed in the deep VVDS region of the 
XMM-LSS field.  We targeted galaxies with high-confidence
redshifts (rank 3 or higher) from the VVDS survey, with higher priority 
given to objects with $i<22.5$. At lower priority we targeted galaxies
to $i<23.5$ and galaxies with lower confidence VVDS redshifts, as well as 
galaxies not observed by VVDS, again with higher priority given to objects 
with $i<22.5$.  

\begin{figure}[t]
\plotone{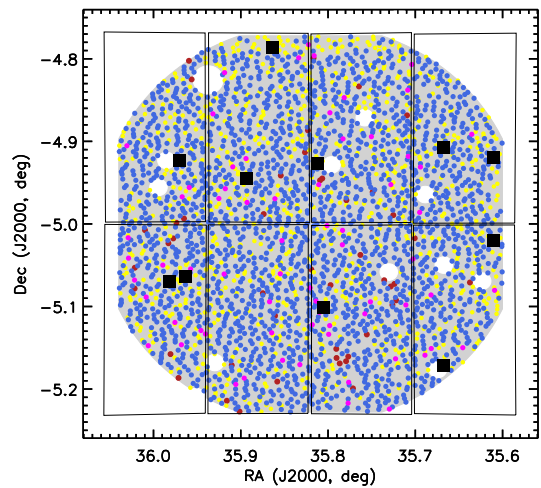}
\caption{\small
Example of the target layout on a PRIMUS slitmask in the XMM-LSS field.  
The eight
rectangular boxes mark the locations of the eight IMACS CCDs.  
The corners of the IMACS field of view is heavily vignetted,
so only the light grey shaded region is used when choosing
targets. Colored points mark the locations of objects assigned slits
on this mask.  High-priority objects (magenta: calibration F stars,
blue: primary galaxy sample) are selected first, and then the
remaining mask area is filled with either lower priority, fainter
galaxies (yellow) or empty sky slits (red).  Black squares show the
locations of alignment stars.}
\label{fig:examplemask}
\end{figure}

\section{Mask Design} \label{sec:mask_design}

\subsection{Overall Strategy}

A single PRIMUS slitmask covers $\sim$0.18 deg$^2$ of sky.  Slitmasks were
designed such that every area of the sky targeted by PRIMUS was
observed with two slitmasks; thus every galaxy has two chances of
being placed on a slitmask.  Within a given field PRIMUS prioritized areas of 
the sky with existing {\it Spitzer} coverage.
Slitmask centers are generally chosen such that
there is very little overlap between adjacent slitmasks, except in the
COSMOS field where the central regions of the field are covered by up
to four or six masks.  Small irregularities in the overall slitmask layout
within a field are due to guide star constraints.
The geometry of the slitmasks and bright star regions 
is tracked using mangle polygons \citep{Hamilton04,Swanson08}.

Using nod-and-shuffle \citep{Glazebrook01}, 
two slits are milled for each target: one for the 
object and one for sky.  We chose to nod in both the RA and Dec directions,
instead of in the Dec direction only, as is usually done.  
In this way, if a bad column
on the CCD affects data in one slit at a given wavelength, it will not
affect the data in the other slit at the same wavelength.  In hindsight,
the horizontal nod led to non-negligible data reduction issues due to 
scattered light within the camera, and is not necessarily recommended.
The first 23 (17\%) of PRIMUS masks designed had slit widths of either 
0.8\arcsec \ or 1.0\arcsec \ and slit lengths of 1.2\arcsec, with nods 
of 1.6\arcsec \ in RA and 3.2\arcsec \ in Dec and a charge shuffle of 8 pixels.
For the bulk of the survey (the remaining 109 masks) these numbers
were increased to a slit width of 1.0\arcsec \ and a slit length of 1.6\arcsec \
to increase the S/N in the data.  The corresponding nods were 2.0\arcsec \ in 
RA and 4.0\arcsec \ in Dec, with a shuffle of 10 pixels.
The corresponding area on the detector allowed for each object is 
two rectangles, each 140 by 20 pixels, shifted with respect to one another
by 10 pixels in the RA direction. (Extracted spectra are 150 pixels in length;
any overlap with a close neighbor will effectively reduce the spectra by 10
pixels at the blue end.)
Figure \ref{fig:zoom} shows a small portion of imaging in one field, 
the corresponding 
slit assignment for one slitmask covering this area (essentially 
a zeroth-order calibration image), and the dispersed raw data.

Slitmasks typically have $\sim$2,500 targets per mask, including 
alignment stars, 
spectrophotometric standard stars, galaxy and AGN targets, and empty sky 
slits.  We avoided regions around bright stars due to difficulties in 
defining targets and due to scattered light.  For space-based target imaging 
in the COSMOS field, we used a smaller bright star mask.
The typical number of primary targets per slitmask is $\sim$1,600 (note that 
objects can be observed on multiple slitmasks).
Individual targets were allowed to be observed on multiple slitmasks, which
often occurred as there are two masks per pointing.  In our science fields,
21\% of the slits were occupied by ``repeat'' observations of isolated 
galaxy and AGN targets that did not have collisions (see below).  

Figure \ref{fig:examplemask} shows the layout of an example PRIMUS 
slitmask in the 
XMM-LSS field.  The grey circular region shows the usable portion of the 
IMACS field of view in f/2 mode; there is strong vignetting in the corners of 
the IMACS camera.  Each of the colored points shows an object for which a 
slit was milled on this mask.  The black squares mark the locations of
alignment stars used to acquire the field. 
High-priority targets are selected first: primary galaxies are shown in blue
and calibration F stars are shown in magenta.  The mask is then filled with
lower-priority galaxies, shown in yellow, and empty sky slits, shown in red,
that are used in our data reduction.  
During mask design, we optimized slit positions for the local sidereal time 
(LST) of observation 
but always designed masks to have a position angle of 0 degrees.

\subsection{Density-Dependent Sampling}
\label{sec:density}
When designing slitmasks one must take care to not have spectra of
adjacent objects overlap on the detectors.  
The footprint of our spectra on the detectors corresponds to 
an area of $30\arcsec \times  8\arcsec$ on the sky, and thus any 
close pairs of galaxies can have only one galaxy of the pair observed 
on a given slitmask. Observing two slitmasks per pointing greatly alleviates
this problem.  However, galaxies to $i=23.5$ are sufficiently clustered 
in the plane of the sky that even with two slitmasks per pointing we would 
undersample the densest regions in a way that would be difficult to recover
without modeling our completeness using simulations.  To alleviate this 
issue, we employ a density-dependent sampling method in which objects with more 
neighbors are sparse-sampled.  This strategy yields a primary sample 
in which each object has on average only one close neighbor 
(within the $30\arcsec \times  8\arcsec$ footprint on the sky).  
As these sampling rates are 
known for every object, we know the appropriate weight of each object in our
primary sample that we do target.  We can therefore recover unbiased 
two-point correlation functions and environmental statistics 
after applying the appropriate weights.

To calculate the targeting weight, we first divide the galaxy sample into 
two groups: group A contains the brighter galaxies to our ``100\% sampling 
limiting magnitude'' (e.g., $i<22.5$) and group B 
contains galaxies 0.5 magnitude fainter (e.g., $22.5<i<23.0$).   
The number of 
potential collisions (objects whose spectra would overlap on the CCD) 
in the joint A+B sample is
calculated for each galaxy in the sample.  Any galaxy in group B
that has more than two collisions is then sparse sampled at a rate of 30\%.  As 
discussed above, this
sparse sampling (as a function of magnitude) 
is done to ensure that the fainter galaxies, which may 
have higher redshift incompleteness, do not dominate the 
assigned slits compared to the brighter galaxies.

After the sample of fainter galaxies has been sparse-sampled, we again 
count the number of potential collisions for every galaxy in the 
surviving A+B set.  
Then a density-dependent sampling probability is assigned, which 
depends on the number of collisions, where 
\begin{equation}
p(N_\mathrm{collisions}) = \left\{
\begin{array}{cc}
1 & N_\mathrm{collisions} \leq 2 \\
\frac{1.8}{N_\mathrm{collisions}} & N_\mathrm{collisions} > 2
\end{array}\right\}
\end{equation}
where $N_{collisions}$ includes a self-count.  Therefore, 
if $N_{collisions}=2$, then both galaxies can be observed on separate slitmasks.
Only if $N_{collisions}>2$ is a density-dependent sampling weight required.

Galaxies are then assigned to slitmasks using these probabilities, where 
every galaxy is given a random number between 0 and 1 such that if the 
random number is smaller than $p(N_{collisions})$, then the target is 
assigned a slit. This sample of surviving density- and sparse-sampled 
objects is our primary sample.
The random sampling is performed for each object, so it is possible that 
some collisions will remain after the subsampling.  We use a numerator 
of 1.8 in equation 7 as a compromise between decreasing the number of 
surviving triplets (and hence improving completeness) and increasing the 
size of the primary sample.  The advantage of the density-dependent a 
priori sampling is that most collisions are removed by a process with a 
well-defined sampling weight.

Using this algorithm we assign slits to 
$\sim$80\% of galaxies to $i<22.5$ (or the equivalent ``100\% sampling limiting
magnitude'' in each field) and 96\% of the density-dependent sample (the
galaxies that passed the density-dependent sampling).  
As the sampling is done using 
well-defined weights which are tracked throughout the target selection 
process and the random seeds are saved, we can apply a ``targeting''
weight to each object in statistical analyses such as the correlation 
function or luminosity function, which will account for the fact that 
not all galaxies could be assigned slits.

\section{Observations} \label{sec:observations}

\begin{figure}[t]
\plotone{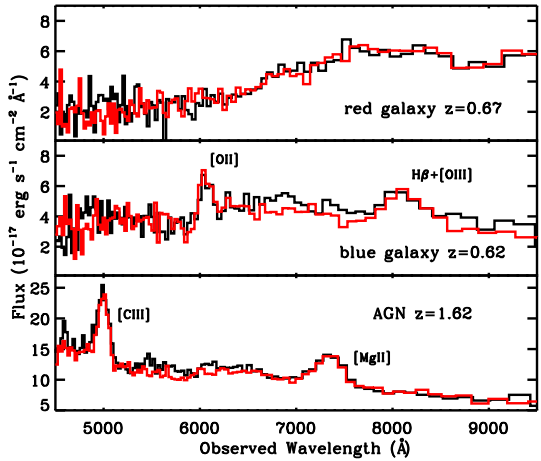}
\caption{\label{fig:spectra}
\small Examples of three PRIMUS spectra, showing an early-type galaxy (top),
a late-type emission line galaxy (middle), and a broad-line AGN (bottom). The
data for each object from the two nod-and-shuffle slits are not combined; here
they are overlaid in black and red.  Prominent emission line features are noted.
}
\end{figure}

The low-dispersion prism was commissioned in January 2006. 
PRIMUS slitmasks were observed during a total of 37 nights in a series of 
observing runs spanning two years, from March 2006 to January 2008, 
in blocks of two to seven nights with time allocated through Arizona, 
MIT, Harvard, Michigan, and NOAO.  A total of 132 slitmasks were 
observed in science fields as well as 10 additional slitmasks in calibration 
fields.

Typical integration times for each slitmask were approximately one
hour, with the exception of masks in the COSMOS field, which were
observed for $\sim$1.5 hours to obtain higher S/N  spectra and to 
target to a slightly fainter limiting magnitude.  
We observed in nod-and-shuffle mode,
with 60 second dwell times, in sets of 16 (8 on, 8 off).
We used the atmospheric dispersion corrector (ADC) on IMACS for all of 
our observations.  Dome flats through the $BVRI$ filters were taken in the
afternoon, as well as twilight flats through the $BRI$ filters.
Masks were aligned with typically two bright stars per CCD, for a total 
of 16 alignment stars.
Alignment was often checked and corrected for, if needed, after half an hour of
exposing.  Arcs were taken during the night at the end of science
exposures for each slitmask to facilitate wavelength calibration.

The seeing varied from FWHM$\sim0.4$\arcsec to $1.4$\arcsec, with typical values
$<1$\arcsec.  The S/N of the
data was monitored in real time by reducing bright objects on a portion of
the mask; masks with low S/N were observed for longer periods of time or 
again on subsequent nights under better conditions.

\section{Merging Photometry} \label{sec:photomerge}

\begin{figure*}[t]
\plottwo{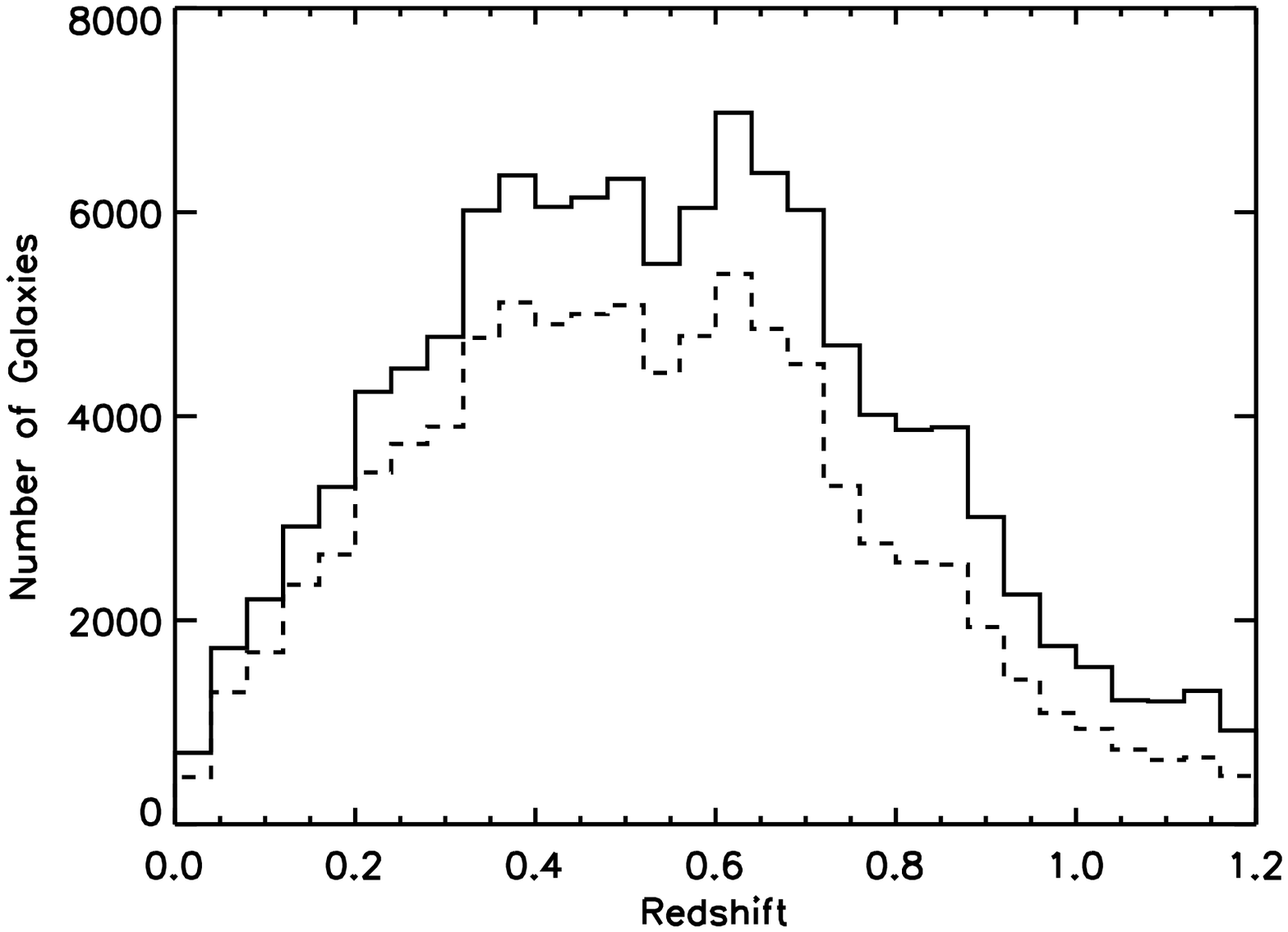}{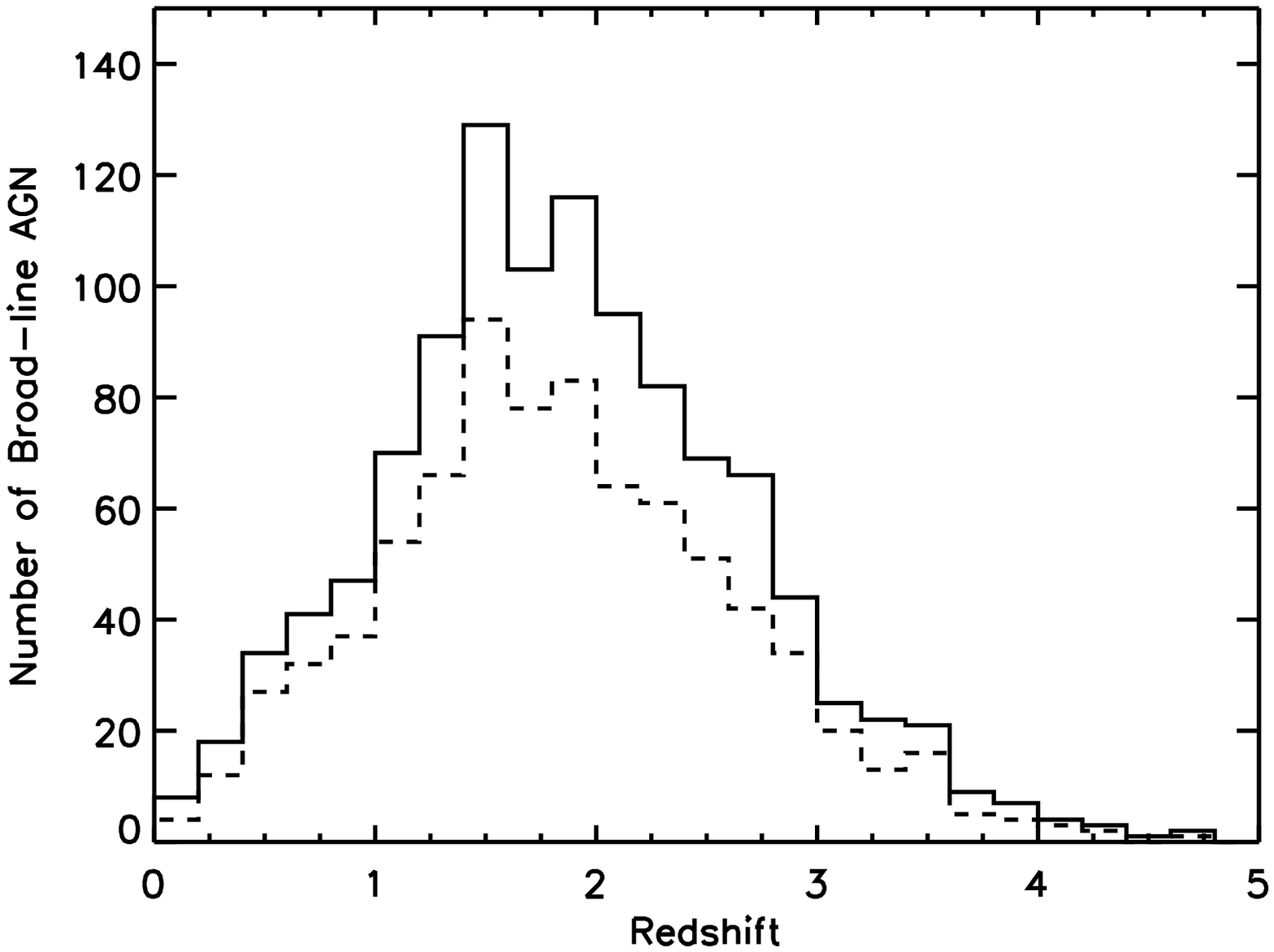}
\caption{\label{fig:zhist}
\small Left: Redshift distribution of galaxies in PRIMUS with robust 
redshifts.  The solid line is for the full sample and the dashed line is
for the primary sample.  Galaxy redshifts are fit from $z=0$ to $z=1.2$,
the range shown here. 
Right: Redshift distribution of broad-line AGN in PRIMUS with robust 
redshifts.  
Broad-line AGN redshifts are fit from $z=0$ to $z=5$,
the range shown here.
}
\end{figure*}

\begin{figure*}[t]
\plotone{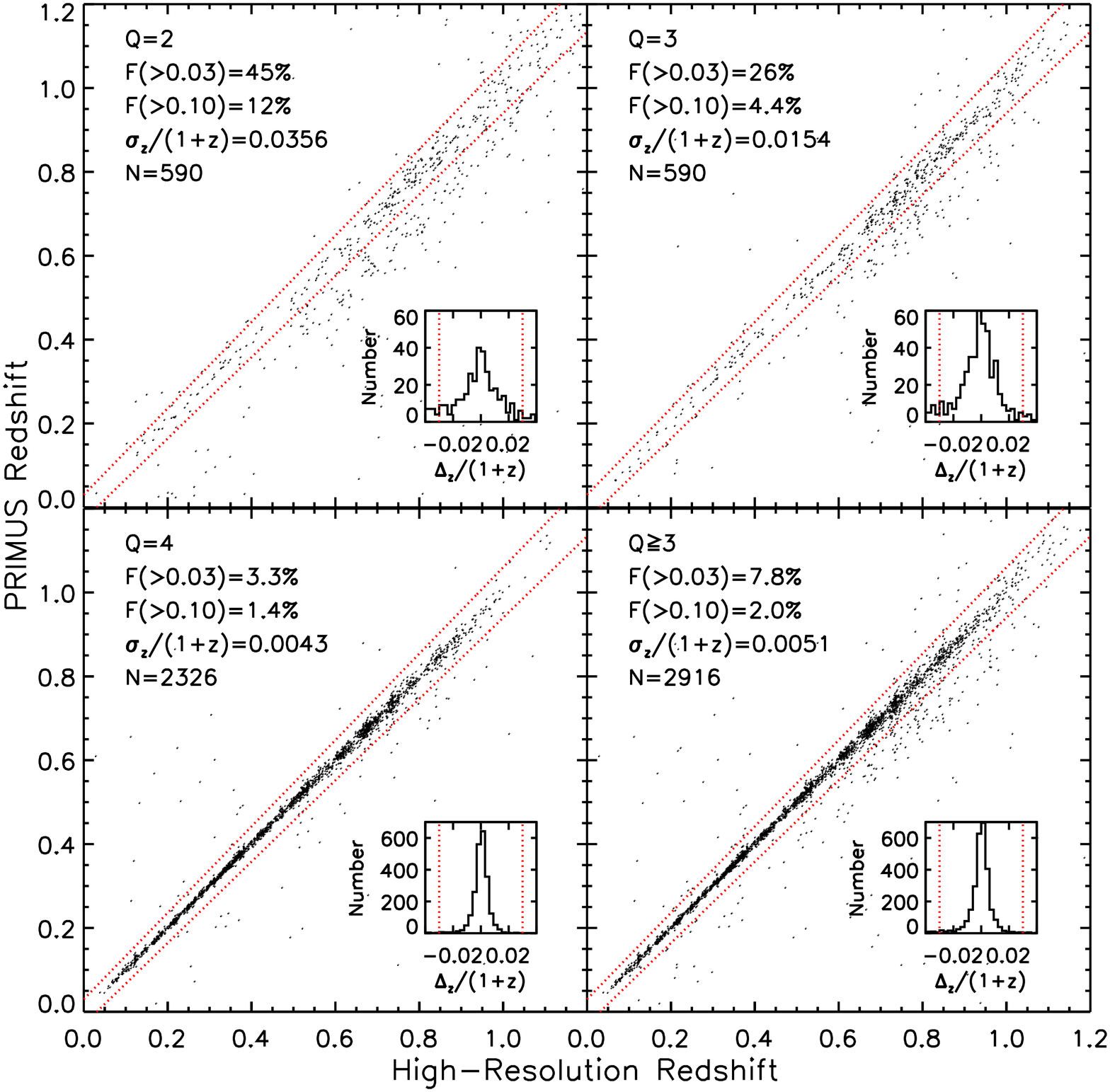}
\caption{\label{fig:zvsz}
\small Comparison of PRIMUS redshifts versus high-resolution spectroscopic
redshifts from either DEEP2, VVDS, or zCOSMOS, for a sample of sources
with $0<z<1.2$ and magnitudes $i<22.5$ (VVDS, zCOSMOS) or $R<22.8$ (DEEP2).
Objects with a PRIMUS redshift confidence of Q$=$2, 3, and 4 are shown in 
the panels in the upper left, upper right, and lower left, while objects 
with Q$\geq$3 are shown in the lower right panel.
Robust redshifts are defined as objects with Q$=$3 or 4.  
The insets show the distribution of $\ddzz$ within 0.5\% for each sample.
Of the sources with $\ddzz<0.03$, the dispersion is 
$\dzz$=0.0154 for Q$=$3, 0.0043 for Q$=$4, and 0.0051 for Q$\geq$3.
The outlier rate of objects in this sample with $\ddzz>0.03$ is 
26\% for Q$=$3, 3\% for Q$=$4, and 8\% for Q$\geq$3, 
while the outlier rate with $\ddzz>0.10$ is 
4\% for Q$=$3, 1\% for Q$=$4, and 2\% for Q$\geq$3.
The outlier rate varies slightly between fields, 
depending on the optical photometric bands available.  
Details are given in Cool et al. (in preparation).  
}
\end{figure*}

Optical ground-based data in each of the PRIMUS fields was gathered not just
for targeting purposes, but to use in fitting for redshifts and in
deriving K-corrections (details are given in Cool et al., in preparation).
When fitting for the redshift, the photometric data points are not
upweighted relative to the spectroscopic data.  Every pixel is weighted
by its corresponding inverse variance, and each photometric band is 
treated as one pixel. The spectroscopic data therefore dominate the 
redshift fit.

Table 4 lists the optical imaging used in each field, in addition to
the telescope and camera that it was taken with, as well as zeropoint
offsets that were applied (details are given below), and 
the reference or source of the data.  We adopt a minimum error floor
to the photometry in each field, between 2\%-10\% in each band, which
is used during redshift determination.  For catalogs that were on the
Vega system, we applied Vega-to-AB corrections as listed in the table.
Catalogs that were on the AB system did not require corrections.
Details of the {\it GALEX}, {\it Spitzer}, and X-ray data in our fields
will be given elsewhere.  

To test the relative calibration of the photometry in each field we
selected stars with good photometry (i.e., unsaturated, high S/N) and
constructed color-color diagrams using every combination of the
available filters.  We then compared the observed position of the
stellar locus (corrected for Galactic reddening; \citealt{Schlegel98,
O'Donnell94}) to the locus predicted by the \citet{Pickles98} stellar
library for metal-poor dwarfs (i.e., similar to stars in the Galactic
halo; \cite{Ivezic08} ).  When synthesizing photometry we used the
best-available filter curves in each field, convolved with the
detector quantum efficiency, telescope throughput, and atmospheric
transmission (see Table 4).

In several of our fields we found non-negligible systematic offsets
between the observed and predicted stellar loci.  To correct for these
offsets, in the CDFS-SWIRE field we applied the recommended zeropoint
corrections from
\cite{Rowan-Robinson08}, in COSMOS we applied the offsets derived by 
\cite{Ilbert09}, and in our XMM-LSS field we used the \cite{Erben09}
recommended offsets.  For the DEEP2 fields we adopted a different
strategy.  First, we retrieved $ugriz$ PSF photometry of F stars
observed by the SDSS in each field and applied the recommended
absolute calibration offsets to the measured magnitudes.\footnote{See
http://howdy.physics.nyu.edu/index.php/Kcorrect for a description of
the absolute AB calibration of the SDSS photometry.}  We only
considered stars with $17<r<19.5$ to ensure good measurements and to
avoid saturation effects.  We then fit the observed photometry
with \cite{Kurucz93} models on a grid of metallicity,
T$_{\mathrm{eff}}$, and $\log(g)$ \citep{Lejeune97} and computed
synthetic $BRI$ magnitudes.  Comparing the observed and synthetic
photometry yielded the offsets listed in Table 4.  Applying
these various corrections to the photometry in each field led to
excellent correspondence between the observed stellar locus, and the
stellar locus predicted by the \citet{Pickles98} stellar library.

\begin{table*}[h]
\tablewidth{0pt}
\label{tab:photo}
\small
\caption{\small Optical photometric imaging and zeropoints used in PRIMUS science and calibration fields.}
\resizebox{\columnwidth}{!} {
	\begin{tabular*}{0.9\columnwidth}{llccccc}
          \colrule
          \colrule
          Band & telescope\footnotemark[a] & $\lambda_{eff}$ & zeropoint & min. & Vega$\rightarrow$AB & ref. \cr
          &        &                 & offset    &  error    &    & \cr
          \colrule
          \colrule
          \multicolumn{6}{c}{CDFS-SWIRE} \cr
          $U$ &  CTIO  &   3672.0  &   0.049  &  0.05 &    0.708 & \footnotemark[b] \cr
          $g$ &  CTIO  &   4733.3  &  -0.034  &  0.05 &   -0.080 & \footnotemark[b] \cr
          $r$ &  CTIO  &   6243.0  &  -0.095  &  0.05 &    0.168 & \footnotemark[b] \cr
          $i$ &  CTIO  &   7613.2  &  -0.010  &  0.05 &    0.393 & \footnotemark[b] \cr
          $z$ &  CTIO  &   8932.3  &  -0.030  &  0.05 &    0.531 & \footnotemark[b] \cr
          \colrule  
          \multicolumn{6}{c}{CDFS-CALIB} \cr
          $U$ &  ESO-2m  &   3647.5  &  0.0  &  0.02 &    0.770 & \footnotemark[c] \cr
          $B$ &  ESO-2m  &   4554.4  &  0.0  &  0.02 &   -0.130 & \footnotemark[c] \cr
          $V$ &  ESO-2m  &   5358.1  &  0.0  &  0.02 &   -0.020 & \footnotemark[c] \cr
          $R$ &  ESO-2m  &   6432.4  &  0.0  &  0.02 &    0.190 & \footnotemark[c] \cr
          $I$ &  ESO-2m  &   8523.6  &  0.0  &  0.02 &    0.490 & \footnotemark[c] \cr
          \colrule  
          \multicolumn{6}{c}{COSMOS} \cr
          $u$ &  CFHT    &   3805.0  &  -0.054  &  0.05 &     0.0 & \footnotemark[d] \cr
          $g$ &  Subaru  &   4728.6  &  -0.024  &  0.02 &     0.0 & \footnotemark[e] \cr
          $r$ &  Subaru  &   6249.1  &  -0.003  &  0.02 &     0.0 & \footnotemark[e] \cr
          $i$ &  CFHT    &   7582.6  &   0.007  &  0.02 &     0.0 & \footnotemark[d] \cr
          $i$ &  Subaru  &   7646.0  &  -0.019  &  0.02 &     0.0 & \footnotemark[e] \cr
          $z$ &  Subaru  &   9011.0  &  0.037  &  0.02 &     0.0 & \footnotemark[e] \cr
          \colrule
          \multicolumn{6}{c}{DEEP2 - 23hr and 02hr} \cr
          $B$ &  CFHT  &    4402.0 &  -0.011  &  0.05 &     0.0 & \footnotemark[f] \cr
          $R$ &  CFHT  &    6595.1 &  -0.034  &  0.05 &     0.0 & \footnotemark[f] \cr
          $I$ &  CFHT  &    8118.7 &   0.035  &  0.05 &     0.0 & \footnotemark[f] \cr
          $i$ &  CFHT  &    7659.1 &   0.015  &  0.05 &     0.0 & \footnotemark[g] \cr
          $z$ &  CFHT  &    8820.9 &   0.025  &  0.05 &     0.0 & \footnotemark[g] \cr
          \colrule
\cr
	\end{tabular*}
}	
\resizebox{\columnwidth}{!} {
	\begin{tabular*}{0.9\columnwidth}{llccccc}
          \colrule
          \colrule
          Band & telescope\footnotemark[a] & $\lambda_{eff}$ & zeropoint & min. & Vega$\rightarrow$AB & ref. \cr
          &        &                 & offset    &  error    &    & \cr
          \colrule
          \colrule
          \multicolumn{6}{c}{DEEP2 - 02hr} \cr
          $u$ &  MMT  &   3604.1  &  -0.087  &  0.05 &     0.0 & \footnotemark[h] \cr
          $g$ &  MMT  &   4763.5  &   0.079  &  0.05 &     0.0 & \footnotemark[h] \cr
          $i$ &  MMT  &   7770.5  &  -0.036  &  0.05 &     0.0 & \footnotemark[h] \cr
          \colrule
          \multicolumn{6}{c}{DLS} \cr
          $B$ &  CTIO  &   4382.0  &    0.0  &  0.05  &   -0.080 & \footnotemark[i] \cr
          $V$ &  CTIO  &   5398.8  &    0.0  &  0.05  &    0.012 & \footnotemark[i] \cr
          $Rc$ &  CTIO  &   6501.5  &    0.0  &  0.05  &    0.212 & \footnotemark[i] \cr
          $z$ &  CTIO  &   8966.4  &    0.0  &  0.05  &    0.000 & \footnotemark[i] \cr
          \colrule
          \multicolumn{6}{c}{ELAIS-S1} \cr
          $B$ &  ESO-2m  &   4344.1  &    0.0  &  0.05  &  -0.090 & \footnotemark[j] \cr
          $V$ &  ESO-2m  &   5455.6  &    0.0  &  0.05  &   0.020 & \footnotemark[j] \cr
          $R$ &  ESO-2m  &   6441.6  &    0.0  &  0.05  &   0.206 & \footnotemark[j] \cr
          $I$ &  VLT     &   8086.3  &    0.0  &  0.10  &   0.456 & \footnotemark[k] \cr
          \colrule
          \multicolumn{6}{c}{SDSS} \cr  
          $u$ &  APO-2.5m  &   3546.0  &  -0.04  &  0.05   &   0.0 & \footnotemark[l] \cr
          $g$ &  APO-2.5m  &   4669.6  &   0.01  &  0.02   &   0.0 & \footnotemark[l] \cr
          $r$ &  APO-2.5m  &   6156.2  &   0.01  &  0.02   &   0.0 & \footnotemark[l] \cr
          $i$ &  APO-2.5m  &   7471.6  &   0.03  &  0.02   &   0.0 & \footnotemark[l] \cr
          $z$ &  APO-2.5m  &   8917.4  &   0.04  &  0.03   &   0.0 & \footnotemark[l] \cr
          \colrule
          \multicolumn{6}{c}{XMM-LSS and VVDS} \cr  
          $u$ &  CFHT  &   3805.0  &  -0.10  &  0.05   &   0.0 & \footnotemark[d] \cr
          $g$ &  CFHT  &   4824.0  &  -0.02  &  0.02   &   0.0 & \footnotemark[d] \cr
          $r$ &  CFHT  &   6199.7  &  -0.05  &  0.02   &   0.0 & \footnotemark[d] \cr
          $i$ &  CFHT  &   7582.6  &  -0.05  &  0.02   &   0.0 & \footnotemark[d] \cr
          $z$ &  CFHT  &   8793.0  &  -0.05  &  0.02   &   0.0 & \footnotemark[d] \cr
          \colrule
	\end{tabular*}
}			

\footnotetext[a]{Cameras used at each telescope: CFHT is Megacam, except for
the $BRI$ imaging in the DEEP2 fields, which is CFHT12k.  CTIO is MOSAIC on 
the Blanco 4m, ESO-2.2m is WFI, MMT is Megacam, Subaru is Suprime-Cam, 
VLT is VIMOS.}
\footnotetext[b]{\cite{Lonsdale03}, http://www.astro.caltech.edu/$\sim$bsiana/cdfs\_opt}
\footnotemark[c]{\cite{Wolf04}}
\footnotemark[d]{CFHTLS T0003 release}
\footnotemark[e]{\cite{Capak07}}
\footnotemark[f]{\cite{Coil04}}
\footnotemark[g]{L. Lin, private communication, lihwailin@asiaa.sinica.edu.tw}
\footnotemark[h]{M. Ashby, private communication, mashby@cfa.harvard.edu}
\footnotemark[i]{\cite{Wittman06}}
\footnotemark[j]{\cite{Berta06}}
\footnotemark[k]{\cite{Berta08}}
\footnotemark[l]{\cite{Aba09}}
\end{table*}

\section{Data Summary} \label{sec:data_summary}

\begin{figure}[t]
\plotone{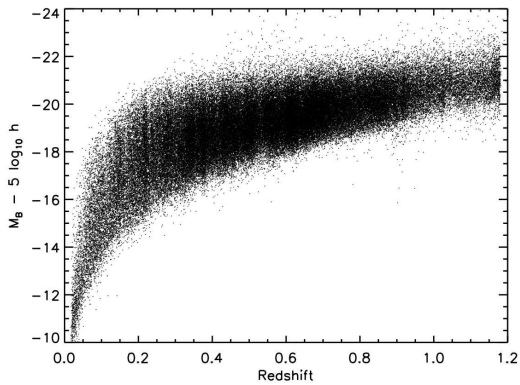}
\caption{\label{fig:absmag}
\small 
Absolute $M_{\rm B} - 5 \ {\rm log} \ h$ magnitude versus redshift for 
120,050 PRIMUS sources (galaxies and AGN) between $0<z<1.2$ across all of our science fields.  
}
\end{figure}

\begin{figure*}[thb]
\plottwo{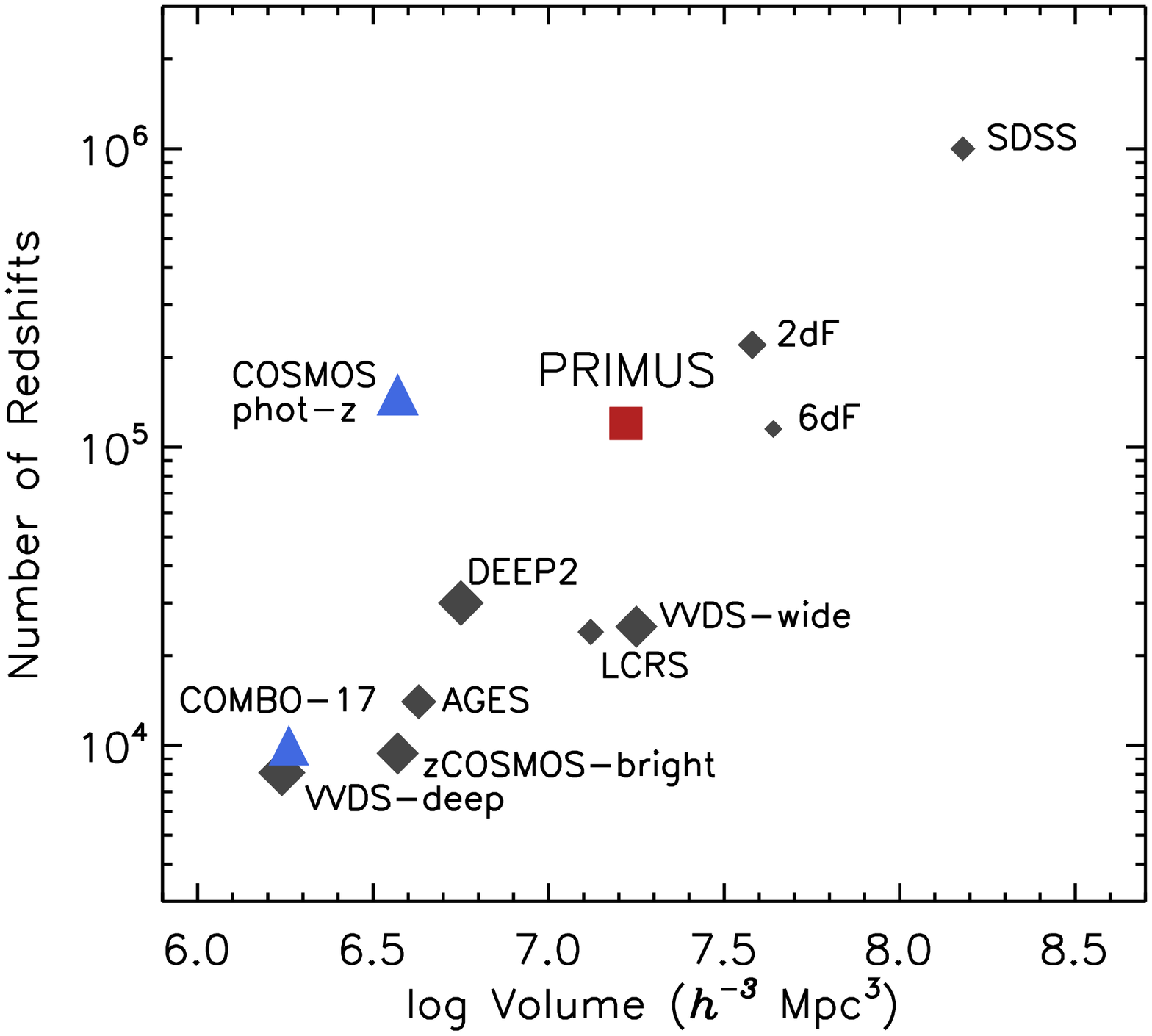}{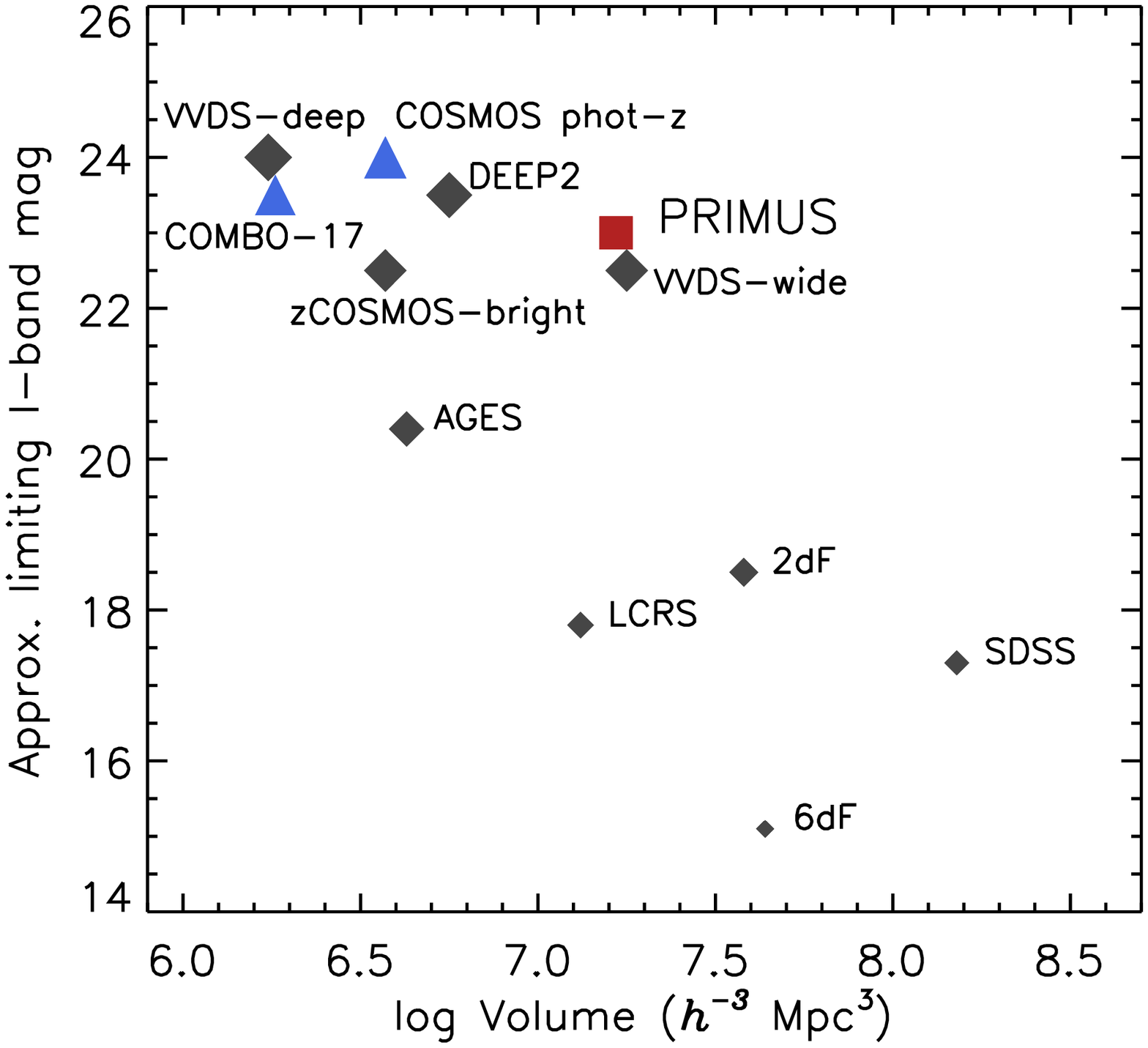}
\caption{\label{fig:surv_compare}
\small Left: Volumes and number of unique robust redshifts for various galaxy
redshift surveys to $z\sim1$.  Low-redshift surveys shown include SDSS, 2dFGRS, 
6dF, and LCRS; the rest are intermediate-redshift surveys.  All are 
spectroscopic redshift surveys except for ``COSMOS phot-z'' \citep{Ilbert09}, 
which are 30-band
photometric redshifts with $\sim$1\% redshift precision, and COMBO-17.  
The symbol size reflects the depth of the survey.
Right: Volumes and approximate {\it I}$_{\rm AB}$ limiting magnitude for the 
same galaxy redshift surveys to $z\sim1$. 
}
\end{figure*}

A summary of the number of objects observed by PRIMUS, and the number
with robust redshifts, is given in Table 1.  PRIMUS observed a total
of 132 slitmasks in our science fields, covering 9.1 deg$^2$, most of
which has multi-wavelength coverage from {\it Spitzer, GALEX, XMM} or
{\it Chandra}.  The fraction of objects to the ``100\% targeting
magnitude limit'' that were observed is very high, $\sim$80\%.  Of the
total $\sim$270,000 objects targeted to $i\sim23.5$, we obtain robust
redshifts for $\sim$130,000 unique sources, $\sim$90,000 of which are
galaxies and AGN in the primary sample.  Our redshift success rate of
$\sim$50\% to our full targeting depth of i$\sim$23.5 is roughly
comparable to spectroscopic surveys of similar depth; in DEEP2 the
redshift success fraction is $\sim$65\% and in the first epoch VVDS
data it is 80\% for ``flag 2, 3, 4 or 9'' sources, which have a
similar redshift quality as our robust ($Q=3+4$; see below) redshifts
here.  The redshift success rate of PRIMUS is a function of S/N of the
spectrum, which closely correlates with galaxy magnitude. At bright
magnitudes ($i<21$) the redshift success rate is $\sim$90\% and it
falls to $\sim$70\% at $i\sim22.5$, with little dependence on 
apparent galaxy color.  A detailed discussion of the
dependence of the redshift success fraction on magnitude and color
will be presented in Cool et al. (in preparation) and 
Moustakas et al. (in preparation).

Example spectra are shown in Figure \ref{fig:spectra}. The resolution
of PRIMUS is high enough to allow us to distinguish broad-line AGN
(lower panel) from emission line galaxies (middle panel) and to
identify continuum features such as the 4000 \AA \ break in early-type
galaxies (top panel).  However, it is
  generally not high enough to measure stellar velocity dispersions,
  emission-line widths, or the strengths of low equivalent width
  nebular (narrow) lines.

The data reduction and redshift fitting pipelines are described in
Cool et al. (in preparation).  Briefly, we determine redshifts
  by fitting the observed spectra and multiwavelength
  photometry simultaneously with an empirical library of galaxy
  spectral templates on a coarse redshift grid in the range
  $0<z<1.2$.  For each object, we construct the redshift probability
  distribution function, $P(z)$, by marginalizing over the galaxy
  templates and identify the first several $P(z)$ maxima.  We then
  refine our search on a finer grid centered on the most likely
  redshift (i.e., the mode of $P(z)$) and compute the final redshift
  and uncertainty, $z_{\rm best}$ and $\sigma_{z}$, as the first and
  second moments of the refined $P(z)$ distribution.  

To assess the
  confidence level of our redshift measurement, we compute a statistic
  $\zeta\propto(\sigma_{z}/(1+z_{\rm best}))/\sqrt{\Delta\chi^{2}}$,
  where $\Delta\chi^{2}$ is the difference between the first and
  second minima (maxima) of the $\chi^{2}(z)$ ($P(z)$) distribution.
  In effect, $\zeta$ is proportional to the 
 narrowness of the $P(z)$ distribution around the first
  maximum and inversely proportional to the relative probability between
 the two most likely redshifts.  Small values of $\zeta$ imply
  a sharply peaked, unimodal $P(z)$ distribution, while a $P(z)$
  distribution with a broad maximum or with a potentially significant
  second maximum result in large values of $\zeta$.  

Using our 
  spectroscopic calibration fields, we correlate $\zeta$ with both the
  cumulative fraction in which PRIMUS measures the ``correct''
  high-resolution redshift (i.e., $|z_{\rm PRIMUS}-z_{\rm calib}|/(1+z_{\rm
    calib})<0.03$) and with the outlier rate, corresponding to 
  the fraction of objects
  with $|z_{\rm PRIMUS}-z_{\rm calib}|/(1+z_{\rm calib})>0.03$ at a
  given $\zeta$.  We then associate three intervals of the continuous
  variable $\zeta$ with our redshift confidence flag, $Q$, such that
  we maximize the fraction of correct redshifts and minimize the
  outlier rate.

We also fit each spectrum with a custom suite of AGN templates in
  the range $0<z<5$ and with stellar templates drawn from the
  \citet{Pickles98} stellar library spanning a wide range of spectral
  types.  We use the differences between the various $\chi^{2}$ minima
  to select among the galaxy, AGN, and stellar templates, with some
  empirically motivated modifications; see Cool et al. (in
  preparation), for details.

Figure \ref{fig:zhist} shows the redshift distributions of PRIMUS
galaxies (left panel) and broad-line AGN (right panel).
In each figure the solid
line shows the redshifts distribution for the full sample and the
dashed line for the primary sample. The mean redshift of the survey is
$z=0.56$.

Figure \ref{fig:zvsz} compares the PRIMUS redshift to a high
resolution spectroscopic redshift for sources from either DEEP2, VVDS,
or zCOSMOS in our calibration fields.  Shown are redshifts for a
sample of galaxies with $0<z<1.2$ and magnitude $R<23.3$ in the DEEP2
fields or $i<23.0$ in the VVDS and zCOSMOS fields.  In order to match
the observed redshift distribution of sources in the full PRIMUS
sample, for this comparison we downweight sources in the DEEP2 fields
so as to create a comparable sample with a similar median redshift as
our survey.  We use only the most robust redshifts from zCOSMOS and
VVDS (flag 3 and 4).  For each redshift confidence flag, Q, we list
the fraction of sources with $\ddzz>0.03$ and with $\ddzz>0.10$, the
redshift accuracy $\dzz$, (where we quote the normalised median
absolute deviation, defined as 1.48 $\times$ median($|\Delta
z|/(1+z)$), and the number of objects with each confidence flag.
Objects with a PRIMUS redshift confidence of Q$=$2 are shown in the
upper left panel; these sources have $\dzz$=0.036 and we do not
consider these sources to have a ``robust'' redshift.  Objects with
Q$=$3 are shown in the upper right; these sources have a robust
redshift but a higher dispersion than the Q$=$4 sources shown in the
lower left.  For most science purposes we will use samples with
Q$\geq$3, shown in the lower right panel.  The Q$=$4 sample is the
purest, with the greatest redshift accuracy and lowest outlier rate,
while the Q$\geq$3 sample is somewhat larger.  The redshift accuracy
of the Q$=$4 sample is $\dzz$=0.0043, while the accuracy of the
Q$\geq$3 sample is $\dzz$=0.0051.  The outlier rate of objects in this
sample with $\ddzz>0.03$ is 8\% for Q$\geq$3 and 3\% for Q$=$4, while
the outlier rate with $\ddzz>0.10$ is 2\% for Q$\geq$3 and 1\% for
Q$=$4.

The outlier rate varies slightly between fields, depending on the
optical photometric bands available, in particular the $u$ band.  For
the vast majority of the sample ($\sim$90\% of our sources), the best
fit redshift did not change if the photometric data was or was not
used.  Including the photometric data helped for only $\sim$10\% of
the objects; these are sources where there is a degeneracy between the
Balmer break and the presence of O[II] 3727 \AA \ in our
low-dispersion spectra.  For these sources the inclusion of
photometric data, $u$ band in particular, helped to break the
degeneracy and the redshift changed by up to $\sim$8\% in $\delta
z/(1+z)$.  Details are given in Cool et al. (in preparation).

As an independent check of the redshift precision, \citet{Wong10} 
identify close galaxy pairs in PRIMUS within a projected separation 
of $r_p<50$ \kpch \ on the plane of the sky and examine the 
redshift differences between the pairs.  They find that $\sim$90\% of 
the pair galaxies are separated from their close neighbor by 
$\Delta z /(1 + z) \leq 0.005$. As almost all galaxies that are this 
close on the plane of the sky 
are physically associated (i.e., in the same dark matter halo), the 
pair galaxies should be at the same redshift modulo peculiar velocities 
along the line of sight.  Therefore the distribution of the redshift 
differences between close pairs is a strong test of our redshift precision, 
completely independent of comparisons against high resolution redshifts.

Figure \ref{fig:absmag} shows the absolute 
$M_{\rm B} - 5 \ {\rm log} \ h$ magnitude versus redshift for 
116,041 PRIMUS sources between $0<z<1.2$.  Restframe magnitudes
are computed using the kcorrect software package \citep{Blanton07}, 
which fits the sum of a set of basis templates at the PRIMUS redshift to the
 broadband optical photometry available in
each field. The basis templates are based on stellar population
synthesis models and are constrained to produce
a non-negative best-fit template, which is used to estimate
the restframe absolute $M_{\rm B} - 5 \ {\rm log} \ h$ magnitude 
from the nearest available observed photometric band.

Figure \ref{fig:surv_compare} compares PRIMUS to other flux-limited
galaxy redshift surveys, in terms of the number of objects, volume,
and approximate 
limiting magnitude of the survey.  At lower redshifts, we compare
to 2dFGRS \citep{Colless01}, 6dF \citep{Jones09}, LCRS \citep{Shectman96},
and SDSS \citep{York00}.  At higher redshifts we compare to AGES,
DEEP2 \citep{Davis03, Davis07}, COMBO-17 \citep{Wolf03, Wolf04}, 
VVDS \citep{Lefevre05, Garilli08}, and zCOSMOS \citep{Lilly07}, 
as well as the COSMOS 30-band photometric redshifts \citep{Ilbert09}.  
In this comparison we use the number of
unique objects with robust redshifts for each survey.  For SDSS we
compare to the main flux-limited galaxy sample and do not
include the luminous red galaxy 
sample.  
For ongoing surveys (VVDS-Wide, zCOSMOS-bright) we show the projected
final numbers with a correction based on the fraction of robust
redshifts (flag 2, 3, 4 or 9) in the current release.
We limit the PRIMUS data to $z<1.2$, which
covers a volume of $V=10^{7.23} \ h^{-3} Mpc^3$.  In the right panel
of Figure \ref{fig:surv_compare} we compare the various redshift
surveys in terms of volume and depth, where we have plotted the {\it
  I}$_{\rm AB}$-band limiting magnitude or an approximate 
equivalent for surveys
limited in other bands.  PRIMUS is the largest-volume, 
intermediate-redshift spectroscopic galaxy redshift 
survey undertaken to date.  The VVDS-wide
survey \citep{Garilli08} is the only other survey covering a
comparable volume at these redshifts, though it is $\sim$half a
magnitude shallower and has a sample size that is 30\% that of PRIMUS.
As we show here, PRIMUS requires a similar amount of total telescope
time as the COSMOS 30-band photometric redshift survey, but PRIMUS 
covers $\sim$5 times as much area with comparable redshift precision.

The PRIMUS survey will allow for the most robust measurements of
galaxy properties and large-scale structure to $z\sim1$ performed to
date.  Initial studies include quantifying triggered star formation in
close galaxy pairs \citep{Wong10}, studying obscured star
formation on the red sequence \citep{Zhu10}, and measuring the luminosity
functions (Cool et al., in preparation) and stellar mass functions 
(Moustakas et al., in preparation) of star-forming and quiescent galaxies. 
 The PRIMUS redshift precision of $\dzz=$0.5\% 
allows for clustering and environment
studies, which integrate over that amount in redshift space due to
peculiar velocities of galaxies within overdense regions.  With PRIMUS
we will be able to measure the galaxy luminosity function, SFR, and
stellar mass density as a function of environment to $z\sim1$ with low
cosmic variance errors.  PRIMUS will also lead to more precise determinations 
of correlation functions for galaxies and AGN at intermediate redshifts 
than has been previously measured, due to the large survey volume spread 
across multiple fields on the sky.

We will present the data reduction, redshift fitting, redshift confidence and 
precision, and survey completeness for PRIMUS in Cool et al. (in preparation), 
to which we refer the reader for additional details about the survey.

\

\acknowledgements

We acknowledge Douglas Finkbeiner, Timothy McKay, Sam Roweis, and Wiphu 
Rujopakarn for their contributions to the PRIMUS
project. We would like to thank the CFHTLS, COSMOS, DLS, and 
SWIRE teams for their public data releases and/or access to early releases.  
We particularly acknowledge Stefano Berta, Carol Lonsdale, 
Brian Siana, Jason Surace, and David Wittman for help with imaging catalogs.
This paper includes data gathered with the 6.5 meter Magellan 
Telescopes located at Las Campanas Observatory, Chile.
We thank the support staff at LCO 
for their help during our observations, and we acknowledge
the use of community access through NOAO observing time.
Some of the data used for this project is from the CFHTLS public data release,
which includes observations obtained with MegaPrime/MegaCam, a joint project 
of CFHT and CEA/DAPNIA, at the Canada-France-Hawaii Telescope (CFHT) which is 
operated by the National Research Council (NRC) of Canada, the Institut 
National des Science de l'Univers of the Centre National de la Recherche 
Scientifique (CNRS) of France, and the University of Hawaii. This work is 
based in part on data products produced at TERAPIX and the Canadian Astronomy 
Data Centre as part of the Canada-France-Hawaii Telescope Legacy Survey, a 
collaborative project of NRC and CNRS.
Funding for PRIMUS has been provided
by NSF grants AST-0607701, 0908246, 0908442,
0908354, and NASA grant 08-ADP08-0019.
RJC is supported by NASA through Hubble Fellowship grant 
HF-01217 awarded by the Space Telescope Science Institute, 
which is operated by the Associated of Universities for
 Research in Astronomy, Inc., for NASA, under contract NAS 5-26555.

\end{document}